\documentclass[11pt]{amsart}
\usepackage{latexsym,amssymb,amsmath,amscd,amsthm}
\numberwithin{equation}{section}
\theoremstyle{remark}
\newtheorem{theorem}{{\bf THEOREM}}[section]

\newcommand{\bq}{\begin{equation}}
\newcommand{\bea}{\begin{array}}
\newcommand{\eea}{\end{array}}

\newcommand{\ga}{\alpha}
\newcommand{\gep}{\epsilon}
\newcommand{\gD}{\Delta}
\newcommand{\gl}{\lambda}
\newcommand{\gL}{\Lambda}
\newcommand{\gb}{\beta}

\newcommand{\mf}{\mathfrak}
\newcommand{\mc}{\mathcal}

\newcommand{\wg}{\wedge}

\newcommand{\go}{\omega}
\newcommand{\gO}{\Omega}
\newcommand{\gG}{\Gamma}

\newcommand{\gs}{\sigma}

\newcommand{\gag}{\gamma}
\newcommand{\gd}{\delta}
\newcommand{\pp}{\partial}

\newcommand{\olra}{\overleftrightarrow}

\newcommand{\tl}{\tilde}
\newcommand{\na}{\nabla}
\newcommand{\gk}{\kappa}

\newcommand{\bs}{\blacksquare}

\newcommand{\bgs}{\bigstar}

\newcommand{\gS}{\Sigma}

\newcommand{{\DDD}}{D\!\!\!\!\!\!-}

\title{REMARKS ON GRAVITY, ENTROPY, AND INFORMATION}

\author{Robert Carroll\\University of Illinois, Urbana, IL 61801}

\date{February, 2006\thanks{email: rcarroll@math.uiuc.edu}}

\begin{document}

\bibliographystyle{plain}

\begin{abstract} 
This is a partially survey collection of material on gravity, entropy, and information
with some new heuristic results related to the WDW equation.
\end{abstract}

\maketitle

\tableofcontents

\section{INTRODUCTION}
\renewcommand{\theequation}{1.\arabic{equation}}
\setcounter{equation}{0}

We gather here some material relating covariant quantum field theory (QFT)
\`a la deDonder-Weyl,
Bohmian mechanics, the WDW equation, differential entropy, and Fisher information.
Some of this is speculative and/or heuristic but the themes suggested seem worth
pursuing.  In particular one sees apparently deep connections between physics
and information theory, which theme was enunciated many years ago by B. Frieden
\cite{f1},
J.A. Wheeler, and others.

\section{DEDONDER-WEYL THEORY}
\renewcommand{\theequation}{2.\arabic{equation}}
\setcounter{equation}{0}

We begin with a sketch of the deDonder-Weyl theory and some applications following
\cite{c1,k3,k7,n1,n2,n3} (cf. also \cite{g1,h7,h23,h73,k18,m2,p6,r2}).  First from
\cite{k3,k7}
we give some information and
discussion of deDonder-Weyl (dDW) theory.  Recall that the
dDW formula for the classical Euler-Lagrange (EL) equations takes the form
\bq\label{10.1}
\frac{\pp p_a^i}{\pp x^i}=-\frac{\pp H}{\pp y^a};\,\,\frac{\pp y^a}{\pp x^i}=\frac
{\pp H}{\pp p_a^i};\,\,p_a^i=\frac{\pp L}{\pp(\pp_iy^a)};\,\,H=p_a^i\pp_iy^a-L
\end{equation}
The (manifest) covariance simply means that space and time variables enter the theory
on a completely equal footing.  The HJ equation of dDW theory is $({\bf 2A})\,\,
\pp_{\mu}S^{\mu}+H(y^a,p_a^{\mu}=\pp S^{\mu}/\pp y^a)=0$ and the 
functional HJ equation of the canonical formalism is
\bq\label{10.2}
\pp_t{\bf S}+{\bf H}\left(y({\bf x}),\pi({\bf x})=\frac{\gd{\bf S}}{\gd y({\bf
x})}\right)=0
\end{equation}
Then in \cite{k7} one writes for $\gS$ a Cauchy surface, $\gS(y=y({\bf x}),\,\,
t=$ constant)
\bq\label{10.3}
{\bf S}=\int_{\gS}(S^{\mu}\go_{\mu})|_{\gS}=\int_{\gS}S^t|_{\gS}
\end{equation}
where $\go_{\mu}=\pp_{\mu}\rfloor (dx^1\wg dx^2\wg dx^3\wg dx^t)$ and $S^t(y^a=
y^a({\bf x}),{\bf x},t)$ is a restriction of the time-like component of the $S^{\mu}(y^a,
x^{\mu})$ to $\gS$.
\\[3mm]\indent
{\bf REMARK 2.1.}
We recall (cf. \cite{l1}) for a vector $X=X^j\pp_j$ one defines
\bq\label{10.4}
X\rfloor f=0;\,\,X\rfloor \go^1=\go^1(X)=<\go^1,X>;\,\,X\rfloor \go_idx^i=
\end{equation}
$$=
(X\rfloor \go_i)dx^i+\go_i(X\rfloor dx^i)=\go_idx^i(X)=\go_iX^i$$
Also the Lie derivative is defined via $({\bf 2B})\,\,{\mf L}_X\go=X\rfloor d\go
+d(X\rfloor \go).\hfill\bs$
\\[3mm]\indent
We recall now that canonical field quantization for say
$L=(1/2)\pp_{\mu}y\pp^{\mu}y-V(y)$ involves $\pi(x)=(\pp L/\pp (\pp_ty(x)))$ with
\bq\label{10.5}
H=\int d^3x(\pp_ty(x)\pi(x)-L)=\int
d^3x\left(\frac{1}{2}\pi^2(x)+(\pp_iy(x))^2+V(x)\right)
\end{equation}
for $\pi(x)=\pp_ty(x)$.  Then one takes $\hat{\pi}(x)=-i\gd/\gd y(x)$ and 
$i\pp_t\psi=\hat{H}\psi$ where 
\bq\label{10.6}
\hat{H}=\frac{1}{2}\int d{\bf x}\left(-\frac{\gd^2}{\gd y^2({\bf x})}+(\pp_iy({\bf x})^2
+V({\bf x})\right)
\end{equation}
\indent
Now going to dDW theory look at
\bq\label{10.7}
\pp_tS^t+\pp_iS^i+\frac{1}{2}\pp_yS^{\mu}\pp_yS_{\mu}+V=0
\end{equation}
and the standard functional HJ equation reads (cf. (2.2))
\bq\label{10.8}
\pp_t{\bf S}+\frac{1}{2}\int d{\bf x}\left[\left(\frac{\gd{\bf S}}{\gd y({\bf x})}
\right)^2+(\pp_iy({\bf x}))^2+2V\right]=0
\end{equation}
From (2.3) one obtains then
\bq\label{10.9}
\pp_t{\bf S}=\int d{\bf x}\pp_tS^t|_{\gS};\,\,\frac{\gd{\bf S}}{\gd y({\bf x})}=
\pp_yS^t|_{\gS}
\end{equation}
The equation for $S^{\mu}|_{\gS}$ can be obtained from the dDW HJ equation by noticing
that when acting on $S^{\mu}|_{\gS}$ the spatial derivative $\pp_i$ turns into the total
derivative $(d/dx^i)=\pp_i+\pp_iy({\bf x})\pp_y$, the last term of which should be
compensated.  Thus the equation for $S^{\mu}|_{\gS}$ assumes the form (signature
$(---+)$)
\bq\label{10.10}
\pp_tS|_{\gS}+\frac{d}{dx^i}S^i|_{\gS}-\pp_iy({\bf x})\pp_yS^i|_{\gS}+\frac{1}{2}
(\pp_yS^t|_{\gS})^2-\frac{1}{2}(\pp_yS^i|_{\gS})^2+V=0
\end{equation}
Substituting $\pp_tS^i|_{\gS}$ from this equation into the right side of (2.9A)
and using (2.9B) one obtains
\bq\label{10.11}
\pp_t{\bf S}+\int d{\bf x}\left(\frac{1}{2}\left(\frac{\gd{\bf S}}{\gd y({\bf
x})}\right)^2+\frac{d}{dx^i}S^i|_{\gS}-\pp_iy({\bf x})\pp_yS^i|_{\gS}-
\frac{1}{2}(\pp_yS^i|_{\gS})^2+V\right)=0
\end{equation}
The second term under the integral does not contribute since it is a total divergence
(this point may need further clarification in some gravitional models).
The third and forth terms together lead to $(1/2)(\pp_iy({\bf x}))^2$ because in
dDW theory $\pp_yS^i=p^i$ and for a scalar field $p^i|_{\gS}=-\pp_iy({\bf x})$.
We have therefore obtained the functional HJ equation (10.8) as a consequence of the
dDW HJ equation (10.7) restricted to the Cauchy surface $\gS$ and a natural hypothesis
(10.3) on relating the HJ eikonal functional ${\bf S}$ to the dWD eikonal functions
$S^{mu}$.

\section{COVARIANT QFT}
\renewcommand{\theequation}{3.\arabic{equation}}
\setcounter{equation}{0}

One shows here following \cite{n1,n2,n3,n4} that the
deterministic evolution of quantum fields is a covariant version of the Bohmian hidden
variable interpretation of quantum field theory (QFT). The deDonder-Weyl (dDW) covariant
canonical formalism is exploited in a novel manner and a covariant Bohmian formulation is
not postulated but derived; this suggests that the Bohmian interpretation could be the
missing link between QM and GR.
The dDW formalism treats space and time variables on an equal footing.  Thus given a 
Lagrangian $L(y^a,\pp_{\mu}y^a,x^{\nu})$ with field variables $y^a$ and $\mu,\nu=1,\cdots,
n)$ one defines polynomials $p_a^{\mu}=\pp L/\pp(\pp_{\mu}y^a)$ and a dDW
Hamiltonian (cf. 2.1))
$H=\pp_{\mu}y^ap_a^{\mu}-L$ such that the Euler-Lagrange (EL) field equations take the
form (cf. (2.1))
\bq\label{8.1}
\pp_{\mu}y^a=\frac{\pp H}{\pp p_a^{\mu}};\,\,\pp_{\mu}p_a^{\mu}=-\frac{\pp H}{\pp y^a}
\end{equation}
The fields are treated as a multitime dDW system evolving in space and time (not just
in time) and everything is manifestly covariant.  Consequently this is an ideal
framework for quantum gravity.  Following now \cite{n1} (cf. also \cite{c1}) one writes
(using only one field $\phi$ for illustration)
\bq\label{8.2}
{\mf A}=\int d^4x{\mf L};\,\,{\mf L}=\frac{1}{2}(\pp^{\mu}\phi)(\pp_{\mu}\phi)-V(\phi);
\,\,\pi^{\mu}=\frac{\pp{\mf L}}{\pp(\pp_{\mu}\phi)}=\pp^{\mu}\phi
\end{equation}
The covariant canonical equations of motion and dDW Hamiltonian (not related to
the energy density) are
\bq\label{8.3}
\pp_{\mu}\phi=\frac{\pp{\mf H}}{\pp\pi^{\mu}};\,\,\pp_{\mu}\pi^{\mu}=-\frac{\pp{\mf H}}
{\pp\phi};\,\,{\mf H}(\pi^a,\phi)=\pi^{\mu}\pp_{\mu}\phi-{\mf L}=\frac{1}{2}\pi^{\mu}
\pi_{\mu}+V
\end{equation}
By introducing the local vector $S^{\mu}(\phi(x),x)$ the dynamics can also be described
by the covariant dDW Hamilton-Jacobi equation and equation of motion
\bq\label{8.4}
{\mf H}\left(\frac{\pp S^a}{\pp\phi},\phi\right)+\pp_{\mu}S^{\mu}=0;\,\,\pp^{\mu}\phi=
\pi^{\mu}=\frac{\pp S^{\mu}}{\pp\phi}
\end{equation}
Note here that $\pp_{\mu}$ acts only on the second argument of $S^{\mu}(\phi(x),x)$
and the corresponding total derivative is $d_{\mu}=\pp_{\mu}+(\pp_{\mu}\phi)
(\pp/\pp\phi)$.  To describe the relation between the covariant HJ equation and the
conventional HJ equation one writes from \eqref{8.3} - \eqref{8.4}
\bq\label{8.5}
\frac{1}{2}\frac{\pp S_{\mu}}{\pp\phi}\frac{\pp S^{\mu}}{\pp\phi}+V+\pp_{\mu}S^{\mu}=0;
\,\,\frac{1}{2}\frac{\pp S_{\mu}}{\pp\phi}\frac{\pp S^{\mu}}{\pp\phi}=\frac{1}{2}
\frac{\pp S^0}{\pp\phi}\frac{\pp S^0}{\pp\phi}+\frac{1}{2}(\pp_i\phi)(\pp^i\phi)
\end{equation}
where $i=1,2,3$ are the space indices and one notes also that $({\bf 3A})\,\,
\pp_{\mu}S^{\mu}=\pp_0S^0+d_iS^i -(\pp_i\phi)(\pp^i\phi)$.
Now introduce the quantity ${\mf S}=\int d^3x S^0$ leading to
\bq\label{8.6}
\frac{\pp S^0(\phi(x),x)}{\pp\phi(x)}=\frac{\gd {\mf S}([\phi(x,t)],t)}{\gd \phi(x,t)};
\,\,\frac{\gd}{\gd \phi(x,t)}=\left.\frac{\gd}{\gd\phi(x)}\right|_{\phi(x)=\phi(x,t)}
\end{equation}
Putting the second equation of (3.5) and (3.6) into the first equation of
(3.5) yields upon integration
then (cf. (2.8))
\bq\label{8.7}
\int d^3x\left[\frac{1}{2}\left(\frac{\gd{\mf S}}{\gd \phi(x,t)}\right)^2+\frac{1}{2}
(\na \phi)^2+V(\phi)\right]+\pp_t{\mf S}=0
\end{equation}
which is the standard non-covariant HJ equation (recall here $\pp^i\phi=
\pp S^i/\pp\phi$ and see Section 2 for a more detailed derivation).
The time evolution of the field 
$\phi(x,t)$ is now given
via $({\bf 3B})\,\,\pp_t\phi(x,t)=\gd{\mf S}/\gd\phi(x,t)$ 
(from the time component in (3.4)) and one
notes that in deriving
\eqref{8.7} it was necessary to use the space part of the equations of motion in 
(3.4); this will be important in the quantum extension below.
\\[3mm]\indent
We recall that QFT can be formulated in the Schr\"odinger picture via
\bq\label{8.8}
\hat{H}\psi=i\hbar\pp_t\psi;\,\,\hat{H}=\int d^3x\left[-\frac{\hbar^2}{2}\left(\frac
{\gd}{\gd\phi(x)}\right)^2+\frac{1}{2}(\na\phi)^2+V(\phi)\right]
\end{equation}
Write now $({\bf 3C})\,\,\psi([\phi(x)],t)={\mf R}([\phi(x)],t)exp(i{\mf
S}([\phi(x)],t)/\hbar)$ and \eqref{8.8} will be equivalent to a set of two real equations
\bq\label{8.9}
\int d^3x\left[\frac{1}{2}\left(\frac{\gd{\mf S}}{\gd\phi(x)}\right)^2+\frac{1}{2}
(\na\phi)^2+V(\phi)+{\mf Q}\right]+\pp_t{\mf S}=0;
\end{equation}
$$\int d^3x\left[\frac{\gd{\mf R}}{\gd\phi(x)}\frac{\gd{\mf S}}{\gd\phi(x)}+{\mf
J}\right] +\pp_t{\mf R}=0;\,\,
{\mf Q}=-\frac{\hbar^2}{2{\mf R}}\frac{\gd^2{\mf R}}{\gd\phi^2(x)};\,\,
{\mf J}=\frac{{\mf R}}{2}\frac{\gd^2{\mf S}}{\gd\phi^2(x)}$$
The second equation is equivalent to
\bq\label{8.10}
\pp_t{\mf R}^2+\int d^3x\frac{\gd}{\gd\phi(x)}\left({\mf R}^2\frac{\gd{\mf
S}}{\gd\phi(x)}\right)=0
\end{equation}
and this represents the unitarity of the theory since it provides a norm $({\bf 3D})\,\,
\int[d\phi(x)]\psi^*\psi=\int [d\phi(x)]{\mf R}^2$ that does not depend on time
(some argument is needed here).
One must also stipulate that the quantity $exp(i{\mf S}/\hbar)$ be single valued.
This formulation also suggests an interesting Bohmian
interpretation stating that the
quantum fields have a deterministic time evolution given by the classical equation
$({\bf 3B})$ and the statistical predictions will be equivalent to those of the
conventional interpretation (cf. \cite{c1,n1} for discussion).  Comparing now \eqref{8.9}
with \eqref{8.7} we see that the quantum field satisfies an equation similar to the
classical one except for the additional nonlocal quantum potential ${\mf Q}$.  There are
no contradictions here with the Bell theory (which specifies local hidden variables) and
the quantum equation of motion will be
\bq\label{8.11}
\pp^{\mu}\pp_{\mu}\phi+\frac{\pp V(\phi)}{\pp\phi}+\frac{\gd Q}{\gd\phi(x,t)}=0
\end{equation}
where $Q=\int d^3x{\mf Q}$.  
\\[3mm]\indent
We will now need a covariant version of the Bohm theory
which goes as follows.  One wants first a quantum version of the classical covariant
dDW HJ equation in \eqref{8.5} and one formulates the classical version first in a
somewhat different way.  Thus let $A([\phi],x)$ be a functional of $\phi$ and a function
of $x$; define then $(\bgs)\,\,dA/d\phi(x)=\int d^4x'(\gd A([\phi],x')/\gd\phi(x))$
where $\gd/\gd \phi(x)$ is a spacetime functional derivative.  If
$A([\phi],x)=A(\phi(x),x)$ (local functional) then $dA([\phi],x)/d\phi(x)=\int d^4x'(\gd
A(\phi(x'),x')/\gd\phi(x))=\pp A(\phi(x),x)/\pp\phi(x)$.  An example of particular
interest here is a functional nonlocal in space but local in time so that
\bq\label{8.12}
\frac{\gd A([\phi],x')}{\gd\phi(x)}=\frac{\gd
A([\phi],x')}{\gd\phi(x,x^0)}\gd(x^{'0}-x^0);\,\,\frac{dA([\phi],x)}{d\phi(x)}=
\end{equation}
$$=\frac{\gd}{\gd\phi(x,x^0)}\int d^3x'A([\phi],x',x^0)$$
One can write the HJ equation in (3.5) as
\bq\label{8.13}
\frac{1}{2}\frac{dS_{\mu}}{d\phi}\frac{dS^{\mu}}{d\phi}+V+\pp_{\mu}S^{\mu}=0
\end{equation}
which is appropriate for the quantum modification.  Similarly the classical equations of
motion in (3.4) can be written as $({\bf 3F})\,\,\pp^{\mu}\phi=dS^{\mu}/d\phi$.
This leads now to the quantum analogue of the classical covariant equation, namely
\bq\label{8.14}
\frac{1}{2}\frac{dS_{\mu}}{d\phi}\frac{dS^{\mu}}{d\phi}+V+{\mf Q}+\pp_{\mu}S^{\mu}=0
\end{equation}
(cf. \cite{p32}).  Here (3.14) is manifestly covariant provided that ${\mf Q}$ in (3.9)
can be written in a covariant form (see below for this).  One can then show that
(3.14) implies (3.9) provided $S^0$ is local in time (so that (3.12) can be used
- cf. (3.6)) and
$S^i$ must be completely local so that $dS^i/d\phi=\pp S^i/\pp\phi$ and hence
$d_iS^i=\pp_iS^i+(\pp_i\phi)(dS^i/d\phi)$ (cf. (3.4)).  Thus in the covariant quantum
theory based on the dDW formalism one must require the validity of $({\bf 3F})$ and
this is nothing but a covariant version of the Bohmian equations of motion written for
an arbitrarily nonlocal $S^{\mu}$.  To produce covariant versions of the remaining 
terms in (3.9) introduce a vector $R^{\mu}([\phi],x)$ which generates a preferred
foliation of spacetime with $R^{\mu}$ normal to the leaves of the foliation.  Then
introduce $({\bf 3G})\,\,{\mf R}([\phi],\gS)=\int_{\gS}d\gS_{\mu}R^{\mu}$
where
$\gS$ is a 3-D leaf generated by $R^{\mu}$.  Similarly a covariant version of
${\mf S}$ is $({\bf 3H})\,\,{\mf S}([\phi],\gS)=\int_{\gS}d\gS_{\mu}S^{\mu}$
with $\gS$ again generated by $R^{\mu}$.  The covariant version of $({\bf 3C})$ is
then $({\bf 3I})\,\,
\psi([\phi],\gS)={\mf R}([\phi],\gS)exp(i{\mf S}([\phi],\gS)/\hbar)$ and for $R^{\mu}$
one postulates the equation
\bq\label{8.15}
\frac{dR^{\mu}}{d\phi}\frac{dS_{\mu}}{d\phi}+{\mf J}+\pp_{\mu}R_{\mu}=0
\end{equation}
In this manner a preferred foliation emerges dynamically as a foliation generated by
the solution $R^{\mu}$ of (3.15) and (3.14).  Note that $R^{\mu}$ plays no classical
role and the existence of a preferred foliation is a purely quantum effect.  Now the
relation betweeen (3.15) and (3.9) is obtained by assuming that nature has chosen a
solution of the form $R^{\mu}=(R^0,0,0,0)$ where $R^0$ is local in time and 
by integration of (3.15) over $d^3x$ with $S^0$ local one sees
that (3.15) is truely a covariant substitute for (3.9).  Finally one has covariant
versions of ${\mf Q}$ and ${\mf J}$ in the form
\bq\label{8.16}
{\mf Q}=-\frac{\hbar^2}{{\mf R}}\frac{\gd^2{\mf R}}{\gd_{\gS}\phi^2(x)};\,\,{\mf J}=
\frac{{\mf R}}{2}\frac{\gd^2{\mf S}}{\gd_{\gS}\phi^2(x)}
\end{equation}
where $\gd/\gd_{\gS}\phi(x)$ is a version of (3.6) in which $\gS$ is generated by 
$R^{\mu}$.  Here $\gS$ depends on $x$ ($x\in\gS$) and $\gS$ is kept fixed in the variation
$\gd_{\gS}\phi(x)$.  Thus (3.14)-(3.15) with (3.16) represent a covariant substitute
for the functional SE (3.8) equivalent to (3.9).  The covariant Bohmian equations
$({\bf 3F})$ imply a covariant version of (3.11), namely
\bq\label{8.17}
\pp^{\mu}\pp_{\mu}\phi+\frac{\pp V}{\pp \phi}+\frac{d{\mf Q}}{d\phi}=0
\end{equation}
Since the last term can also be written as $\gd(\int d^4x{\mf Q}/\gd\phi(x)$ the
equation of motion (3.17) can be obtained by varying the quantum action $({\bf 3J})\,\,
{\mf A}_Q=\int d^4x{\mf L}_Q=\int d^4x({\mf L}-{\mf Q})$.  To summarize one can say that
the convenentional SE corresponds to a special class of solutions of the covariant
canonical quantization of fields given by (3.14), (3.15), and (3.16) for which
$R^i=0,\,\,S^i$ is local, and $R^0,\,S^0$ are local in time.
\\[3mm]\indent
Generalizations are included in \cite{n1} dealing with a larger number of fields and
curved spacetimes.  We indicate some of the equations and refer to \cite{n1} for
discussion.  Thus let $\phi(x)=\{\phi_a(x)\}$ be a collection of fields with action
\bq\label{8.18}
{\mf L}=\frac{1}{2}G^{ab}(\phi,x)g^{\mu\nu}(x)(\pp_{\mu}\phi_a)(\pp_{\nu}\phi_b)+
F^{a\mu}(\phi,x)\pp_{\mu}\phi_a-V(\phi,x)
\end{equation}
In particular $G^{ab},\, F^{a\mu}$, and V are proportional to $|g|^{1/2}$ for convenience
in calculations etc. so $|g|^{1/2}$ is included in the definition of ${\mf L}$.
One writes $G_{ab}G^{bc}=\gd_a^c$ and 
since $G^{ab}\sim |g|^{1/2}$ one notes that if
$\pp_{\mu}\phi_a$ is a tensor then $\pp^{\mu}\phi^a$ is a tensor density.  The canonical
momenta are $\pi^{a\mu}=\pp{\mf L}/\pp(\pp_{\mu}\phi_a)=\pp^{\mu}\phi^a+F^{a\mu}$ and
the dDW Hamiltonian is
\bq\label{8.19}
{\mf H}=\pi^{ab}\pp_{\mu}\phi_a-{\mf L}=\frac{1}{2}(\pp^{\mu}\phi^a)(\pp_{\mu}\phi_a)+V=
\end{equation}
$$=\frac{1}{2}\pi^{a\mu}\pi_{a\mu}-\pi^{a\mu}F_{a\mu}+\frac{1}{2}F^{a\mu}F_{a\mu}+V$$
The corresponding covariant canonical equations of motion are then
\bq\label{8.20}
\pp_{\mu}\phi_a=\frac{\pp{\mf
H}}{\pp\pi^{a\mu}}=\pi_{a\mu}-F_{a\mu};\,\,\pp^{\mu}\pi_{a\mu}=-G_{ab}\frac{\pp{\mf H}}
{\pp\phi_b}\equiv-\pp_a{\mf H}
\end{equation}
Here $\pp_a=G_{ab}\pp^b\ne \pp^bG_{ab}$ (since $G_{ab}$ depends on $\phi$) and the
covariant HJ equations are
\bq\label{8.21}
\pi^{a\mu}=\frac{\pp S^{\mu}}{\pp\phi_a}\equiv\pp^aS^{\mu};
\end{equation}
$$\frac{1}{2}
(\pp^aS^{\mu})(\pp_aS_{\mu})-F_{a\mu}\pp^aS^{\mu}+\frac{1}{2}F^{a\mu}F_{a\mu}
+V+\pp_{\mu}S^{\mu}=0$$
The total derivative is $d_{\mu}=\pp_{\mu}+(\pp_{\mu}\phi_a)\pp^a$ and one shows
explicitly that (3.21) is covariant (cf. \cite{n1}).  The general covariant
generalization of $(\bgs)$ depends on the tensor nature of A and here A is a vector
density $A^{\mu}$ so one writes
\bq\label{8.22}
\frac{dA^{\mu}([\phi],x)}{d\phi(x)}\equiv\frac{e^{\mu}_{\bar{\ga}}(x)}{|g(x)|^{1/2}}
\int d^4x'e^{\bar{\ga}}_{\nu}(x')\frac{\gd A^{\nu}([\phi],x')}{\gd\phi(x)}
\end{equation}
where $e^{\mu}_{\bar{\ga}}$ is the tetrad satisfying $e^{\mu}_{\bar{\ga}}e^{\bar{\ga}\nu}
=g^{\mu\nu}$ ($\bar{\ga}$ is an index in the $SO(1,3)$ group).  Now in (3.21)
one replaces the derivative $\pp^a$ with $d^a=d/d\phi_a$ and adds the Q term where
\bq\label{8.23}
{\mf Q}=-\frac{\hbar^2}{2{\mf
R}}\frac{\gd}{\gd_{\gS}\phi_a}G_{ab}\frac{\gd}{\gd_{\gS}\phi_b} {\mf R}
\end{equation}
Then (3.15) generalizes to
\bq\label{8.24}
(d^aR^{\mu})(d_aS_{\mu})-F_{a\mu}d^aR^{\mu}+{\mf J}+\pp_{\mu}R^{\mu}=0;\,\,
{\mf J}=\frac{{\mf R}}{2}\frac{\gd}{\gd_{\gS}\phi_a}\left(G_{ab}\frac{\gd{\mf
S}}{\gd_{\gS}
\phi_b}-F_{a\mu}r^{\mu}\right)
\end{equation}
where $r^{\mu}=R^{\mu}/(R^{\gl}R_{\gl})^{1/2}$.  The orderings in (3.23) are chosen
to lead to a SE with a Hermitian Hamiltonian.  One uses now the manifestly covariant
forms
\bq\label{8.25}
{\mf R}([\phi],\gS)=\int_{\gS}d\gS_{\mu}\tl{R}^{\mu};\,\,{\mf S}([\phi],\gS)=\int_{\gS}
d\gS_{\mu}\tl{S}^{\mu}
\end{equation}
where $\tl{S}^{\mu}$ and $\tl{R}^{\mu}=R^{\mu}/|g|^{1/2}$ are vectors.  The Bohmian
equations of motion 
\bq\label{8.26}
\pp_{\mu}\phi_a=d_aS_{\mu}-F_{a\mu}
\end{equation}
are then equivalent to the equations obtained by varying the quantum action 
$({\bf 3J})$ and we refer to \cite{n1} for details and further generalization.
\\[3mm]\indent
In \cite{n1}, in addition to the calculations involving $G_{ab}$ (cf. (3.18)-(3.26))
one discusses quantum gravity as follows.  The classical gravitational action is
${\mf A}=\int d^4x|g|^{1/2}R$ where R is the scalar curvature and to write the Lagrangian
in a form appropriate for a canonical treatment one sets
\bq\label{8.33}
|g|^{1/2}R=\frac{1}{2}G^{\ga\gb\mu\gag\gd\nu}(\pp_{\mu}g_{\ga\gb})(\pp_{\nu}g_{\gag\gd})
+total\,\, derivative
\end{equation}
The total derivative term is ignored and one assumes that $G^{\ga\gb\mu\gag\gd\nu}$
and its inverse depend on $g_{\ga\gb}$ but not on its derivatives (cf. \cite{h73,p63}).
The fields $\phi_a$ are the components $g_{\ga\gb}$ of the metric and all 10 components
will be quantized (in contrast to the convential noncovariant canonical quantization
where only the space components are quantized).  One finds then (as before) the following
quantum equations
\bq\label{8.34}
\frac{1}{2}G_{\ga\gb\mu\gag\gd\nu}\frac{dS^{\mu}}{dg_{\ga\gb}}
\frac{dS^{\nu}}{dg_{\gag\gd}}
+{\mf Q}+\pp_{\mu}S^{\mu}=0;
\,\,
{\mf Q}=-\frac{\hbar^2}{2{\mf R}}\frac{\gd}{\gd_{\gS}g_{\ga\gb}}G^r_{\ga\gb\gag\gd}
\frac{\gd}{\gd_{\gS}g_{\gag\gd}}{\mf R};
\end{equation}
$$G_{\ga\gb\mu\gag\gd\nu}\frac{dR^{\mu}}{dg_{\ga\gb}}\frac{d
S^{\nu}}{dg_{\gag\gd}}+{\mf J}+\pp_{\mu}R^{\mu}=0;\,\,{\mf J}=\frac{{\mf
R}}{2}\frac{\gd}
{\gd_{\gS}g_{\ga\gb}}G^r_{\ga\gb\gag\gd}\frac{\gd}{\gd_{\gS}g_{\gag\gd}}{\mf S}$$ where
$G^r_{\ga\gb\gag\gd}=G_{\ga\gb\mu\gag\gd\nu}r^{\mu}r^{\nu}$ (recall
$r^{\mu}=R^{\mu}/(R^{\gl}R_{\gl})^{1/2}$).  The Bohmian equations of motion
$\pp_{\mu}g_{\ga\gb}=G_{\ga\gb\mu\gag\gd\nu}(dS^{\mu}/dg_{\gag\gd})$ are equivalent to
the equations of motion obtained via the quantum action ${\mf A}_Q=\int
d^4x(|g|^{1/2}R-Q)$ and this leads to the equation of motion
\bq\label{8.35}
R^{\mu\nu}-\frac{g^{\mu\nu}}{2}R+|g|^{-1/2}\frac{d{\mf Q}}{dg_{\mu\nu}}=0
\end{equation}
The potential ${\mf Q}$ is a scalar density so one can write ${\mf
Q}=|g|^{1/2}\tl{{\mf Q}}$ where
$\tl{{\mf Q}}$ is a scalar and \eqref{8.35} becomes
\bq\label{8.36}
R^{\mu\nu}+\frac{d\tl{Q}}{dg_{\mu\nu}}-\frac{g^{\mu\nu}}{2}(R-\tl{{\mf Q}})=0
\end{equation}
Another suggestive form is
\bq\label{8.37}
\frac{g^{\mu\nu}}{2}R-R^{\mu\nu}=8\pi G_NT^{\mu\nu};\,\,T^{\mu\nu}=\frac{1}{16\pi G_N}
\left(2\frac{d\tl{{\mf Q}}}{dg_{\mu\nu}}+g^{\mu\nu}\tl{{\mf Q}}\right)
\end{equation}
Note that \eqref{8.37} implies that the Bohmian equations of motion are fully covariant.
By contrast if the quantization of gravity is based on the conventional canonical 
WDW equation that does not treat space and time on an equal footing then the Bohmian
interpretation leads to an equation similar to \eqref{8.37} but with a non-covariant
energy-momentum tensor of the form $T^j\propto d\tl{{\mf Q}}/dg_{ij}$ and
$T^{0\mu}\propto
\tl{{\mf Q}}g^{0\mu}$.  One recalls also that the WDW quantization corresponds to the
case in which $R^i=0$ and $S^i$ is local while $S^0,\,R^0$ are functionals local in time.
\\[3mm]\indent
{\bf REMARK 3.1.}
In \cite{n3} the problem of time in quantum gravity is addressed by weakening the 
Hamiltonian constraint $\hat{H}=0$ to $<\psi|\hat{H}|\psi>=0$ which is consistent with the
classical Hamiltonian constraint.  This can be written as (we shift $g\to h$ here
in thinking of applications below to the deWitt metric and ${}^3h\sim h$)
\bq\label{8.27}
\int{\mf D}h\psi^*\hat{H}\psi=0
\end{equation}
and for $\psi=Rexp(iS/\hbar)$, $\hat{H}\psi=i\hbar\pp_t\psi$ and a stipulation
$(d/dt)\int {\mf D}h\psi^*\psi=0$ one finds that (3.32) holds if $\pp_tS=0$
(note $R\sim {\mf R}$ and $S\sim{\mf S}$ here). 
Hence
the (weak) Hamiltonian constraint (3.32) is consistent with
$\psi=R(h,t)exp(iS(h)/\hbar)$ (implies seems too strong here).
The point here is to allow $i\hbar\pp_t\psi=\hat{H}\psi$ but insist that this not
contradict $\hat{H}=0$ in the classical limit.
Consider then $H=\tl{G}_{AB}(h)\pi^A\pi^B+V(h)$ ($h=\{h_A\}$ and
$\tl{G}_{AB}=\tl{G}_{BA}$) or explicitly
\bq\label{8.28}
\tl{G}_{AB}\pi^A\pi^B\equiv\gk\int
d^3x\tl{G}_{ijk\ell}\pi^{ij}\pi^{k\ell};\,\,V=-\gk^{-1}\int d^3x\sqrt{|h|}{}^3R
\end{equation}
where $\gk=8\pi G$ and
\bq\label{8.29}
\tl{G}_{ijk\ell}=\frac{\sqrt{|h|}}{2}(h_{ik}h_{j\ell}+h_{jk}h_{i\ell}-h_{ij}h_{k\ell})
\end{equation}
($\tl{G}_{ijk\ell}$ differs from $G_{ijk\ell}$ by a factor of $\sqrt{h}$
and this can be absorbed in ${\mf D}h$ as needed yielding $\tl{{\mf D}}h$).
In the quantum case $\pi^A$ becomes $\hat{\pi}^A=-i\hbar(\gd/\gd h_A)\equiv
-i\hbar\pp^A$ and different orderings of the $\hat{\pi}^A$ in $\hat{H}$ become
important.  Some argument shows that a form
$({\bf 3K})\,\,\hat{H}=\hat{\pi}^A\tl{G}_{AB}\hat{\pi}^B+V$ implies
$<\psi|\psi>$ as well as all $<\psi_1|\psi_2>$ are time independent since
\bq\label{8.30}
\frac{d}{dt}\int \tl{{\mf D}}h\psi_1^*\psi_2=\hbar\int \tl{{\mf
D}}h\pp^A[\tl{G}_{AB}(\psi_1^*i
\olra{\pp}^B\psi_2]
\end{equation}
which vanishes because the integral over a total derivative vanishes
(thus unitary time evolution implies the sandwich ordering).
Moreover for
$\hbar\to 0$ (with $c=1$) one obtains densities
\bq\label{8.31}
\tl{G}_{AB}\pp^AS\pp^BS+V=0;\,\,\pp_tR^2+\pp^A[2R^2\tl{G}_{AB}\pp^BS]=0
\end{equation}
which is the classical HJ equation (via $\pi^A=\pp^AS$) and 
\bq\label{8.32}
\dot{h}_A=\pp_th_A=\frac{\pp H}{\pp\pi^A}=2\tl{G}_{AB}\pi^B;\,\,\pp_t\rho+
\pp^A(\rho\dot{h}_A)=0\,\,(\sim\,\frac{d\rho}{dt}=0)
\end{equation}
for $\rho=R^2$.  Hence in fact the conventional strong form of the Hamiltonian
constraint (leading to $\pp_t\rho=0$) does not have the correct classical limit, but
the weaker form does. 
$\hfill\bs$

\section{EXACT UNCERTAINTY AND WDW}
\renewcommand{\theequation}{4.\arabic{equation}}
\setcounter{equation}{0}

In \cite{c15} we sketched some new heuristic results concerning WDW and exact
uncertainty following \cite{c1,h1,h2,h3,r1}.

\subsection{EXACT UNCERTAINTY}

Basically following e.g. \cite{h1,h3} one defines
Fisher information via $({\bf 4A})\,\,F_x=\int dx P(x)[\pp_xlog(P(x))]^2$ and a
Fisher length by $\gd x=F_x^{-1/2}$ where $P(x)$ is a probability density for a
1-D observable x.  The Cramer-Rao inequality says $Var(x)\geq F_x^{-1}$ or simply
$\gD x\geq\gd x$.  For a quantum situation with $P(x)=|\psi(x)|^2$ and $\psi$
satisfying a SE one finds immediatly
\bq\label{1.1}
F_X=\int
dx|\psi|^2\left[\frac{\psi'}{\psi}+\frac{\bar{\psi}'}{\bar{\psi}}\right]^2dx=
\end{equation}
$$=4\int dx\bar{\psi}'\psi'+\int dx|\psi|^2\left[\frac{\psi'}{\psi}-\frac
{\bar{\psi}'}{\bar{\psi}}\right]^2
=\frac{4}{\hbar^2}\left[<p^2>_{\psi}-<p_{cl}^2>_{\psi}\right]$$
where $p_{cl}=(\hbar/2i)[(\psi'/\psi)-(\bar{\psi}'/\bar{\psi})]$ is the classical
momentum observable conjugate to x ($\sim S_X$ for $\psi=Rexp(iS/\hbar)$).
Setting now $p=p_{cl}+p_{nc}$ one obtains after some calculation
$({\bf 4B})\,\,F_x=(4/\hbar^2)(\gD p_{nc})^2=1/(\gd x)^2\Rightarrow \gd
x\gD p_{nc}=\hbar/2$ as a relation between nonclassicality and Fisher
information. Note $<p>_{\psi}=<p_{cl}>_{\psi}$,
$\pp_t|\psi|^2+\pp_x[|\psi|^2m^{-1}p_{cl}]=0$ from the SE, and $(\gD x)(\gD
p)\geq (\gd x)(\gD p)\geq (\gd x)(\gD p_{nc})$.
\\[3mm]\indent
We recall also that from \eqref{1.1} $F_x$ is proportional to the difference of
a quantum and a classical kinetic energy.  Thus $(\hbar^2/4)F_x(1/2m)=(1/2m)
<p^2>_{\psi}-(1/2m)<p^2_{cl}>_{\psi}$ and $E_F=(\hbar^2/8m)F_x$ is added to
$E_{cl}$ to get $E_{quant}$.  By deBroglie-Bohm (dBB) theory there is a quantum
potential
\bq\label{1.2}
Q=\frac{\hbar^2}{8m}\left[\left(\frac{P'}{P}\right)^2-2\frac{P''}{P}\right];\,\,
P=|\psi|^2
\end{equation}
and evidently $({\bf 4C})\,\,<Q>_{\psi}=\int PQdx=(\hbar^2/8m)F_x$ (upon neglecting
the boundary integral term at $\pm\infty$ - i.e. $P'\to 0$ at $\pm\infty$).
\\[3mm]\indent
Now the exact uncertainty principle (cf. \cite{h1,h3,r1}) looks at momentum
fluctuations $({\bf 4D})\,\,p=\na S+f$ with $<f>=\bar{f}=0$ and replaces a 
classical ensemble energy $<E>_{cl}$ by ($P\sim|\psi|^2$)
\bq\label{1.3}
<E>=\int dx P\left[(2m)^{-1}\overline{|\na S+f|^2}+V\right]=<E>_{cl}+
\int dx P\frac{\overline {f\cdot f}}{2m}
\end{equation}
Upon making an assumption of the form $({\bf 4E})\,\,\overline{f\cdot f}=
\ga(x,P,S,\na P,\na S,\cdots)$ one looks at a modified Hamiltonian
$({\bf 4F})\,\,\tl{H}_q[P,S]=\tl{H}_{cl}+\int dx P(\ga/2m)$.  Then,
assuming
\begin{enumerate}
\item
Causality - i.e. $\ga$ depends only on $S,P$ and their first derivatives
\item
Independence for fluctuations of noninteracting uncorrelated ensembles
\item
$f\to L^Tf$ for invertible linear coordinate transformations $x\to L^{-1}x$
\item
Exact uncertainty - i.e. $\ga=\overline{f\cdot f}$ is determined solely by 
uncertainty in position
\end{enumerate}
one arrives at
\bq\label{1.4}
\tl{H}_q=\tl{H}_{cl}+c\int dx\frac{\na P\cdot\na P}{2m P}
\end{equation}
and putting $\hbar=2\sqrt{c}$ with $\psi=\sqrt{P}exp(iS/\hbar)$ a SE is obtained.
\\[3mm]\indent
As pointed out in \cite{c2} in the SE situation with Q as in \eqref{1.2}, in
3-D one has
\bq\label{1.5}
\int PQd^3x\sim -\frac{\hbar^2}{8m}\int\left[2\gD P-\frac{1}{P}(\na
P)^2\right]d^3x=\frac{\hbar^2}{8m}\int\frac{1}{P}(\na P)^2d^3x
\end{equation}
since $\int_{\gO}\gD Pd^3x=\int_{\pp\gO}\na P\cdot{\bf n}d\gS$ can be assumed zero for
$\na P=0$ on $\pp\gO$. 
Hence 
\\[3mm]\indent
We recall now (cf. \cite{b2,c1,c8}) that the relation between the SE and the quantum
potential (QP) is not 1-1.  The QP Q depends on the wave function $\psi=Rexp(iS/\hbar)$
via $Q=-(\hbar^2/2)(\gD R/R)$ for the SE and thus the solution of a quantum HJ equation,
involving S and R(via Q), requires the companion ``continuity" equation to determine
S and R (and thence $\psi$).  There is some lack of uniqueness since Q determines R only
up to uniqueness for solutions of $\gD R+(2m/\hbar^2)QR=0$ and even then the HJ equation
$S_t+\cdots=0$ could introduce still another arbitrary function (cf. \cite{c1,c8}).
\begin{theorem}
Given that any quantum potential for the SE has the form \eqref{1.2} (with
$\na P=0$ on $\pp\gO$) it follows
that the quantization can be identified with momentum fluctuations of the type
studied in \cite{h3} and thus has information content as described by the Fisher
information.  
Thus we see that given a SE described via a probability distribution
$P\,\,(=|\psi|^2)$ one can identify this equation as a quantum model arising from a
classical Hamiltonian $\tl{H}_{cl}$ perturbed by a Fisher information term as in (4.4).
Thus the quantization involves an information content with entropy significance
(cf. here \cite{c2,o1}) for entropy connections).  This suggests that any quantization of
$\tl{H}_{cl}$ arises (or can arise) through momentum perturbations related to Fisher
information and it also suggests that $P=|\psi|^2$ (with $\int Pd^3x=1$) should be 
deemed a requirement for any solution $\psi$ of the related SE (note $\int Pd^3x
=1$ eliminates many putative counterexamples).  Thus once P is specified as a
probability distribution for a wave function $\psi=\sqrt{P}exp(iS/\hbar)$ arising
from a SE corresponding to a quantization of $\tl{H}_{cl}$, then Q can be expressed
via Fisher information.  Similarly given Q as a Fisher information perturbation
of $\tl{H}_{cl}$ (arising from momentum fluctuations involving P as in (4.4)) there is a
unique wave function $\psi=\sqrt{P}exp(iS/\hbar)$ satisfying the corresponding SE.
\end{theorem}

\subsection{WDW}

The same sort of arguments can be applied for the WDW equation following \cite
{h1,h2,p2,r1,s1} (cf. also \cite{a2,d1,g3,k1,k17,k18,m2,p3,p4,r4,r5,s3,s5,s6,s8,s9,s11} 
for WDW). Thus take
an ADM situation
\bq\label{2.1}
ds^2=-(N^2-h^{ij}N_iN_j)+2N_idx^idt+h_{ij}dx^idx^j
\end{equation}
and assume dynamics generated by an action $({\bf 4G})\,\,A=\int dt[\tl{H}+\int
{\mf D}hP\pp_tS]$.  One will have equations of motion $({\bf 4H})\,\,
\pp_tP=\gd\tl{H}/\gd S$ and $\pp_tS=-\gd\tl{H}/\gd P$ (cf. \cite{c1,h2}).  A
suitable ``classical" Hamiltonian is 
\bq\label{2.2}
\tl{H}_c[P,S]=
\int{\mf D}hPH_0\left[h_{ij},\frac{\gd S}{\gd h_{ij}}\right];
\end{equation}
$$H_0=
\int dx\left[N\left(\frac{1}{2}G_{ijk\ell}\pi^{ij}\pi^{k\ell}+V(h_{ij})\right)-
2N_i\na_j\pi^{ij}\right]$$
where $G_{ijk\ell}$ is the deWitt (super)metric $({\bf 4I})\,\,
G_{ijk\ell}=(1/\sqrt{h})(h_{ik}h_{j\ell}+h_{i\ell}h_{jk}-h_{ij}h_{k\ell})$ and
$V\sim \hat{c}\sqrt{h}(2\gL-{}^3R)$.  Then thinking of $\pi^{ij}=\gd S/\gd
h_{ij}+ f^{ij}$ and e.g. $\tl{H}_q=\tl{H}_c+(1/2)\int{\mf D}hP\int dx
NG_{ijk\ell}
\overline{f^{ij}f^{k\ell}}$ one arrives via exact uncertainty at a Fisher information
contribution (cf. \cite{c15,f1,h1,h2})
\bq\label{2.3}
\tl{H}_q[P,S]=\tl{H}_{cl}+\frac{c}{2}\int{\mf D}h\int dx
NG_{ijk\ell}\frac{1}{P}\frac {\gd P}{\gd h_{ij}}\frac{\gd P}{\gd h_{k\ell}}
\sim \tl{H}_{cl}+\frac{c}{2}\int{\mf D}hN{\mf Q}
\end{equation}
with $\hbar=2\sqrt{c}$ and $\psi=\sqrt{P}exp(iS/\hbar)$ resulting in
(for $N=1$ and $N_i=0$)
\bq\label{2.4}
\left[-\frac{\hbar^2}{2}\frac{\gd}{\gd h_{ij}}G_{ijk\ell}\frac{\gd}{\gd
h_{k\ell}}+V\right]\psi=0
\end{equation}
with a sandwich ordering ($G_{ijk\ell}$ in the middle - cf. also Section 3 and
\cite{k18}).  In general there are also constraints 
\bq\label{2.5}
\frac{\gd \psi}{\gd N}=\frac{\gd \psi}{\gd
N_i}=\pp_t\psi=0;\,\,\na_j\left(\frac{\gd \psi}{\gd h_{ij}}\right)=0
\end{equation}
We note here (keeping $N=1$ with $N_i=0$)
\bq\label{2.6}
\frac{\gd}{\gd h_{ij}}\left(G_{ijk\ell}\frac{\gd}{\gd
h_{k\ell}}\sqrt{P}e^{iS/\hbar}\right)=
\left[\frac{\gd G_{ijk\ell}}{\gd h_{ij}}\left(
\frac{1}{2}P^{-1/2}\frac{\gd P}{\gd h_{k\ell}}+\frac{iP^{1/2}}{\hbar}
\frac{\gd S}{\gd h_{k\ell}}\right)+\right.
\end{equation}
$$+G_{ijk\ell}\left\{-\frac{1}{4}P^{-3/2}\frac{\gd P}{\gd h_{k\ell}}\frac
{\gd P}{\gd h_{ij}}+\frac{1}{2}P^{-1/2}\frac{\gd^2P}{\gd h_{k\ell}\gd h_{ij}}-
\frac{P^{1/2}}{\hbar^2}\frac{\gd S}{\gd h_{k\ell}}\frac{\gd S}{\gd
h_{ij}}+\right.$$
$$\left.\left.+\frac{i}{2\hbar}P^{-1/2}\left(\frac{\gd P}{\gd
h_{k\ell}}\frac{\gd S}{\gd h_{ij}}+\frac{\gd S}{\gd h_{k\ell}}\frac{\gd P}{\gd
h_{ij}}\right) +\frac{iP^{1/2}}{\hbar}\frac{\gd^2S}{\gd h_{k\ell}\gd
h_{ij}}\right\}\right]e^{iS/\hbar}$$
Therefore writing out the WDW equation gives (cf. \cite{c15})
\bq\label{2.7}
-\frac{\hbar^2}{4P}\frac{\gd}{\gd h_{ij}}\left[G_{ijk\ell}\frac{\gd P}{\gd
h_{k\ell}}\right]+
\end{equation}
$$+\frac{\hbar^2}{8P^2}G_{ijk\ell}\frac{\gd P}{\gd h_{k\ell}}
\frac{\gd P}{\gd h_{ij}}+G_{ijk\ell}\left[
\frac{\hbar^2}{8P}\frac{\gd^2P}
{\gd h_{ij}\gd h_{ij}}+\frac{1}{2}\frac{\gd S}{\gd h_{k\ell}}
\frac{\gd S}{\gd h_{ij}}\right]+V=0;$$
$$2P\frac{\gd G}{\gd h_{ij}}\frac{\gd S}{\gd
h_{k\ell}}+G\left(\frac{\gd P}{\gd h_{k\ell}}\frac{\gd S}{\gd h_{ij}}+
\frac{\gd S}{\gd h_{k\ell}}\frac{\gd P}{\gd h_{ij}}\right)+2PG\frac
{\gd^2S}{\gd h_{k\ell}\gd h_{ij}}=0$$
\indent
It is useful here to compare with $-(\hbar^2/2m)\psi''+V\psi=0$ which for
$\psi=Rexp(iS/\hbar)$ yields
\bq\label{2.8}
\frac{1}{2m}S_x^2+V+Q=0;\,\,Q=-\frac{\hbar^2}{4m}\frac{R''}{R}=\frac{\hbar^2}{8m}
\left[\frac{2P''}{P}-\left(\frac{P'}{P}\right)^2\right]
\end{equation}
along with $\pp(R^2S')=\pp(PS')=0$ (leading to (4.5)).  The analogues here are then in
particular
\bq\label{2.9}
\frac{1}{2m}S_x^2\sim \frac{1}{2}G_{ijk\ell}\frac{\gd S}{\gd h_{k\ell}}\frac{\gd
S}{\gd
h_{ij}};\,\,Q=\frac{\hbar^2}{8m}\left[\frac{2P''}{P}-\left(\frac{P'}{P}\right)^2
\right]\sim
\end{equation}
$$\sim-\frac{\hbar^2}{4P}\frac{\gd}{\gd h_{ij}}\left[G_{ijk\ell}\frac{\gd P}{\gd
h_{k\ell}}\right]+G_{ijk\ell}\left\{\frac{\hbar^2}{8P^2}\frac{\gd P}{\gd
h_{k\ell}}
\frac{\gd P}{\gd h_{ij}}+\frac{\hbar^2}{4P}\frac{\gd^2P}{\gd h_{ij}\gd
h_{k\ell}}\right\}$$
We note that the Q term arises directly from
\bq\label{2.10}
Q=-\frac{\hbar^2}{2}P^{-1/2}\frac{\gd}{\gd h_{ij}}\left(G_{ijk\ell}\frac{\gd P^{1/2}}
{\gd h_{k\ell}}\right)
\end{equation}
and corresponds really to a density with
\bq\label{2.11}
\int {\mf D}h\,PQ=-\frac{\hbar^2}{2}\int {\mf D}h P^{1/2}\frac{\gd}{\gd h_{ij}}
\left(G_{ijk\ell}\frac{\gd P^{1/2}}{\gd h_{k\ell}}\right)
\end{equation}
But from $\int {\mf D}h\gd[\,\,\,]=0$ one has (cf. (4.3))
\bq\label{2.12}
\int{\mf D}h P^{1/2}\frac{\gd}{\gd h_{ij}}\left(G_{ijk\ell}\frac{\gd P^{1/2}}{\gd
h_{k\ell}}\right)=-\int{\mf D}h\frac{\gd P^{1/2}}{\gd h_{ij}}G_{ijk\ell}\frac
{\gd P^{1/2}}{\gd h_{k\ell}}
\end{equation}
\begin{theorem}
Given a WDW equation of the form (4.9) with associated quantum potential given via
(4.15) (or (4.16)) it
follows that the quantum potential gives rise to momentum fluctuations 
of Fisher information type as in 
(4.8) (for $N=1$).
Thus let us assume there exists a suitable ${\mf D}f$ as in Section 4.3 below which
is a measure in the (super)space of fields $h$.  Then there is an integration by parts
formula (4.20) which removes the need for considering surface terms in 
integrals $\int d^3x$ (cf. \cite{d1} for cautionary remarks about Green's theorem,
etc.).  Consequently given a WDW equation of the form (4.9) with corresponding Q as in
(4.15) (and
$\psi=
\sqrt{P}exp(iS/\hbar)$, one can show that the equation can be modelled on a perturbation
of a classical $\tl{H}_c$ via a Fisher information type perturbation as in (4.8)
(cf. here \cite{c1,c2,c15,f1,h2}).  Here P represents a probability density of fields
$h_{ij}$ which determine $G_{ijk\ell}$ (and V incidentally) and the very existence of a 
quantum equation (i.e. WDW) seems to require entropy type input via Fisher information
fluctuation of fields.  This suggests that quantum gravity requires a statistical
spacetime (an idea that has appeared before - cf. \cite{c1}).
\end{theorem}
\indent
{\bf REMARK 4.1.}
We note from \cite{a2,f3,f50,g2,w2} that the ``superspace" $=Riem/Diff$ with the
deWitt metric
$G_{ijk\ell}=G_{k\ell ij}$ is a collection of manifolds called a stratified manifold
and therefore the
calculations involving ${\mf D}h$ here (as well as in \cite{h1,h2})
must be appropriately determined.$\hfill\bs$

\subsection{SOME FUNCTIONAL CALCULUS}

We go here to \cite{b1,c1,h2,h13} and will first sketch the derivation of (3.4)
following \cite{h1,h2} (cf. also \cite{c1}).  The relevant functional calculus goes as
follows.  One defines a functional F of fields $f$ and sets
\bq\label{4.1}
\gd F=F[f+\gd f]-F[f]=\int dx\frac{\gd F}{\gd f_x}\gd f_x
\end{equation}
Here e.g. $dx\sim d^4x$ and in the space of fields there is assumed to be a measure
${\mf D}f$ such that $\int {\mf D}f\equiv\int {\mf D}f'$ for $f'=f+h$ (cf. \cite{b1,h2}).
Then evidently $({\bf 4J})\,\,\int {\mf D}f(\gd F/\gd f)=0$ when $\int
{\mf D}f\,F[f]<\infty$.  Indeed 
\bq\label{4.2}
0=\int {\mf D}f(F[f+\gd f]-F[f])=\int dx\gd f_x\left(\int{\mf D}f\frac{\gd F}{\gd
f_x}\right)
\end{equation}
and this provides an integration by parts formula
\bq\label{4.3}
\int {\mf D}f\,P\left(\frac{\gd F}{\gd f}\right)=-\int {\mf D}f\,\left(\frac{\gd P}{\gd
f}\right)F
\end{equation}
for $P[f]$ a probability density functional.  Classically a probability density functional
arises in discussing an ensemble of fields and conservation of probability requires
\bq\label{4.4}
\pp_tP+\sum_a\int dx\frac{\gd}{\gd f_x^a}\left.\left(P\frac{\gd H}{\gd g_x^a}\right|_
{g=\gd S/\gd f}\right)
\end{equation}
where $g_x^a$ is the momentum corresponding to $f_x^a$; thus one assumes a motion equation
\bq\label{4.5}
\pp_tS+H\left(f,\frac{\gd S}{\gd f},t\right)=0
\end{equation}
The equations of motion here are then
\bq\label{4.6}
\pp_tP=\frac{\gD\tl{H}}{\gD S};\,\,\pp_tS=-\frac{\gD\tl{H}}{\gD P}
\end{equation}
where $({\bf 4K})\,\,\tl{H}(P,S,t)=<H>=\int{\mf D}f PH(f,(\gd S/\gd f),t)$.
The variational theory here involves functionals $I[F]=\int {\mf D} f\,\xi(F,\gd F/\gd f)$
and one can write
\bq\label{4.7}
\gD I=I[F+\gD F]-I[F]=\int{\mf D}f\left[\frac{\pp\xi}{\pp F}\gD F+\int dx\left(\frac
{\pp\xi}{\pp(\gd F/\gd f_x)}\right)\frac{\gd (\gD F)}{\gd f_x}\right]=
\end{equation}
$$=\int{\mf D} f\left[\frac{\pp\xi}{\pp F}-\int dx\frac{\gd}{\gd f_x}\left(\frac
{\pp\xi}{\pp(\gd F/\gd f_x)}\right)\right]\gD F+$$
$$+\int dx\int{\mf D}f\frac{\gd}{\gd f_x}\left[\left(\frac{\pp\xi}{\pp(\gd F/\gd
f_x}\right)\gd F\right]$$
Assuming the term $\int{\mf D}f[\,\,\,]\gD F$ is finite the last integral vanishes and
one obtains $({\bf 4L})\,\,\gD I=\int {\mf D}f(\gD I/\gD F)\gD F$, thus defining a
variational derivative
\bq\label{4.8}
\frac{\gD I}{\gD F}=\frac{\pp\xi}{\pp F}-\int dx\frac{\gd}{\gd f_x}\left(\frac
{\pp\xi}{\pp(\gd F/\gd f_x)}\right)
\end{equation}
In the Hamiltonian theory one can work with a generating function S such that
$({\bf 4M})\,\,g=\gd S/\gd f$ and $\pp_tS+H(f,\gd S/\gd f,t)=0$ (HJ equation) and
solving this is equivalent to $\pp_tf=\gd H/\gd g$ and $\pp_tg=-\gd H/\gd f$ (cf. 
\cite{h2}).  Once S is specified the momentum density $g$ is determinied via 
$g=\gd S/\gd f$ and an ensemble of fields is specified by a probability density functional
$P[f]$ (and not by a phase space density functional $\rho[f,g]$.  In the HJ formulation
one writes $({\bf 4N})\,\,V_x[f]=\pp f_x/\pp t=(\gd H/\gd g)|_{g=
\gd S/\gd f)}$ and hence the associated continuity equation $\pp_t\int {\mf D}fP$ is
\bq\label{4.9}
\pp_tP+\int dx\frac{\gd}{\gd f_x}[PV_x]=0
\end{equation}
provided $<V_x>$ is finite.
\\[3mm]\indent
Now after proving (4.4) one proceeds as follows to produce a SE.  The Hamiltonian
formulation gives $({\bf 4O})\,\,\pp_tP=\gD\tl{H}/\gD S$ and 
$\pp_tS=-\gD\tl{H}/\gd P$ where the ensemble Hamiltonian is
\bq\label{4.10}
\tl{H}=\tl{H}[P,S,t]=<H>=\int{\mf D}f PH[f,\gd S/\gd f,t]
\end{equation}
where P and S are conjugate variables.  The equations $({\bf 4O})$
arise from $\gD\tl{A}=0$ where $\tl{A}=\int dt[-\tl{H}+\int {\mf D}fS\pp_tP$.  One
specializes here to quadratic Hamiltonian functions
\bq\label{4.11}
H_c[f,g,t]=\sum_{a,b}dx K_x^{ab}[f]g_x^ag_x^b+V[f]
\end{equation}
and to this is added a term as in (4.4) to get $\tl{H}$ (which does not depend on S).
Hence from $({\bf 4O})$ with $\pp_tf_x=\gd H_c/\gd g_x$ one obtains
following (4.26)
\bq\label{4.12}
\pp_tP+\int dx\frac{\gd}{\gd f_x}\left[P\frac{\gd H}{\gd g_x}\right]_{g=\gd S/\gd f}=0
\end{equation}
(cf. 4.25)).  The other term in $\tl{H}$ is simply 
\bq\label{4.13}
(\hbar^2/4)\int {\mf D}f\int PK_x^{ab}(\gd P/\gd f_x^a)(\gd P/\gd f_x^b)(1/P^2)
\end{equation}
and
this provides a contribution to the HJ equation via $\pp_tS=-\gD\tl{H}/\gD P$ which will
have the form
\bq\label{4.14}
Q=-\frac{\hbar^2}{4}P^{-1/2}\int dx\frac{\gd}{\gd f_x^a}\left(K_x^{ab}\frac{\gd
P^{1/2}} {\gd f_x^b}\right)\sim \int dx{\mf Q}
\end{equation}
corresponding to (4.15).  We note further then from (4.17)
\bq\label{4.15}
Q\sim \frac{\hbar^2}{2}\int dx G_{ijk\ell}\frac{\gd P^{1/2}}{\gd h_{ij}}\frac{\gd P^{1/2}}
{\gd h_{k\ell}}\sim\frac{\hbar^2}{8}\int dx G_{ijk\ell}\frac{1}{P}\frac{\gd P}
{\gd h_{ij}}\frac{\gd P}{\gd h_{k\ell}}
\end{equation}
as in (4.8).  Hence Theorem 4.2 is established under the hypotheses indicated concerning
${\mf D}f$ etc.  Some care is needed here in distinguishing densities ${\mf Q}$ from Q;
in view of $\int dx\int{\mf D}h=\int{\mf D}h\int dx$ one can move terms around rather
freely.

\section{REMARKS ON ENTROPY}
\renewcommand{\theequation}{5.\arabic{equation}}
\setcounter{equation}{0}

One recalls (cf. \cite{c1,c2,g7}) that with the SE (under certain circumstances) one has a
differential entropy
${\mf S}=-\int dx \rho log(\rho)$ (1-D for simplicity here) with $\pp_t\rho=-\pp(v\rho)$
and $v=-u=-D\pp log(\rho)$ (diffusion current) leading to 
\bq\label{7.4}
\pp_t{\mf S}=-\int dx\rho_t(log(\rho)+1)=\int dx\left(log(\rho)+1\right)\pp(v\rho)=
\end{equation}
$$=-\int \pp\rho D\pp log(\rho)=D\int\frac{(\pp\rho)^2}{\rho}$$
Thus the Fisher information is the time derivative of the differential entropy and 
there should be some analogue of this for WDW.  There is not a priori a natural
time evolution for WDW but Section 3 provides a way around this.  In any case one might
look for a formula of the form
\bq\label{7.5}
\gd\int {\mf D}hF(S,P,h_{ij})=\int {\mf D}h\left[\frac{\gd F}{\gd S}\gd
S+\cdots\right]=\int{\mf D}h\frac{\gd P^{1/2}}{\gd h_{ij}}G_{ijk\ell}\frac {\gd
P^{1/2}}{\gd h_{k\ell}}
\end{equation}
where F represents some kind of entropy term.  Note 
from \cite{h2} that $f^{ij}\sim(1/P)(\gd P/\gd h_{ij})=\gd log(P)/\gd h_{ij}$ is
claimed to be inconsistent with $\overline{f}^{ij}=0$, but for $<f^{ij}>=
\overline{f}^{ij}=\int {\mf D}hf^{ij}$ we get $\int {\mf D}h(\gd log(P)/\gd h_{ij})=0$
automatically.
Hence
referring now to Section 3, in particular (3.9) - (3.10) and Remark 3.1, one thinks
of ${\mf R}^2=\rho\,\,(=P)$ and looks at (3.31) (with Q and J added).  The second
equation is in fact fixed by the sandwich ordering as
\bq\label{7.6}
\pp_t\rho+\pp^A[2\rho \tl{G}_{AB}\pp^BS]=0
\end{equation}
where $\pp^BS=-i\hbar(\gd/\gd h_B)S$.  Now recall from \cite{c1} that in a Brownian
motion situation the use of a drift velocity $u=D\na log(\rho)=-v=-(1/m)\na S$ is natural
($D=\hbar/2m$).  Another context involving the SE with statistical geometry and a Weyl
space produces a Weyl vector $\phi_i=-\pp_ilog(\rho)$ related to an osmotic velocity
field.  Thus a relation ${\bf u}=-c\phi=c\na log(\rho)$ can be envisioned with $\rho=P\sim
{\mf R}^2$ so that, instead of dealing with $\gd S/\gd h_{ij}=\pi^{ij}-(1/P)
(\gd P/\gd h_{ij})$ one is motivated to consider 
\bq\label{7.7}
\frac{\gd S}{\gd h_B}\sim-\frac{\hat{c}}{P}\frac{\gd P}{\gd h_B}
\end{equation}
provided one is only interested in metric fluctuations (there is no particle mass here
to impede this).  In this case on could work with (5.3) as ($(-i\hbar)^2=-\hbar^2$) 
\bq\label{7.8}
\pp_tP-\frac{\gd}{\gd h_A}\left[2P\tl{G}_{AB}\frac{\hbar^2\hat{c}}{P}\frac{\gd P}{\gd
h_B}\right]=0
\end{equation}
Then for a differential entropy defined via $({\bf 5A})\,\,{\mc S}=-(1/\hat{c})
\int dx\int\tl{{\mf D}}h Plog(P)$ one would have 
\bq\label{7.9}
{\mc S}_t\sim -\frac{\hbar^2}{\hat{c}}\int dx\int\tl{{\mf
D}}hP_t[1+log(P)]=
\end{equation}
$$=-\hbar^2\int dx\int\tl{{\mf D}}h[1+log(P)]\left[\frac{\gd} {\gd
h_A}\left(2\tl{G}_{AB}\frac{\gd P}{\gd h_B}\right)\right]=$$
$$=\hbar^2\int dx\int \tl{{\mf D}}h\frac{2}{P}\tl{G}_{AB}\frac{\gd P}{\gd h_B}\frac{\gd
P}{\gd h_A}\sim 16\int{\mf D}h\int dx P{\mf Q}\sim 16\int\tl{{\mf D}}h\,PQ$$
(cf. (4.16), (4.17), and (4.32)).  One arrives then at a heuristic result
\begin{theorem}
Given the weak constraint situation of
Remark 3.1 
and assuming only metric fluctuations satisfying (5.4) one can define a differential
entropy
$({\bf 5A})$ and express the Fisher information (expressed via the quantum potential Q)
as a time derivative
$\pp_t{\mc S}$.  A similar theorem holds using the covariant Bohmian formulation
with $R^i=0,\,\,S^i$ local etc.
\end{theorem}
\indent
{\bf REMARK 5.1.}
In \cite{b3} one develops a theory of a time direction hidden in quantum mechanics
based on $Q(t)>0$  where Q is the quantum potential.  The idea is that
$\int^tQ(\tau)d\tau$ is a monotone increasing function of time which can be useful
to characterize the direction of time.  We will not go into the idea of a knowledge
functional ${\mc K}$ here except to remark that ${\mc K}\sim Q$ (up to a factor
of $\hbar^2/2$).  In any event this also seems to be compatible with entropy change
as envisioned in (5.1) and Theorem 5.1.$\hfill\bs$

\subsection{ENTROPY AND THE EINSTEIN EQUATIONS}

In \cite{p1} (cf. also \cite{c2,c7,p31,p32,p63,p7}) one takes an entropy functional
($u^a=\bar{x}^a-x^a$ is a perturbation)
\bq\label{01.1} 
S=\frac{1}{8\pi}\int d^4x\sqrt{g}\left[M^{abcd}\na_au_b\na_cu_d+N_{ab}u^au^b\right]
\end{equation}
Extremizing with respect to $u_b$ leads to ($N_{ab}u^au^b=N^{ab}u_au_b$)
\bq\label{01.2}
\na_a\left(M^{abcd}\na_c\right)u_d=N^{bd}u_d
\end{equation}
Note $\int d^4x\sqrt{-g}f\na_au_b=-\int d^4x\sqrt{-g}u_b\na_af$ since via \cite{a5}
one can write
$\gd\sqrt{-g}=-(1/2)\sqrt{-g}g_{\mu\nu}\gd g^{\mu\nu}$ and $\na_ag^{\mu\nu}=0$.
Choosing M and N such that \eqref{01.2} (for all $u_d$) implies the Einstein
equations entails 
\bq\label{01.3}
M^{abcd}=g^{ad}g^{bc}-g^{ab}g^{cd};\,\,N_{ab}= 8\pi\left(T_{ab}-\frac{1}{2}g_{ab}T\right)
\end{equation}
Consequently S becomes
\bq\label{01.4}
S=\frac{1}{8\pi}\int d^4x\sqrt{-g}\left[(\na_au^b)(\na_bu^a)-(\na_bu^b)^2+N_{ab}u^au^b\right]=
\end{equation}
$$=\frac{1}{8\pi}\int
d^4x\sqrt{-g}\left[Tr(J^2)-(Tr(J))^2+8\pi\left(T_{ab}-\frac{1}{2}g_{ab}T\right)u^au^b\right]$$
where $J_a^b=\na_au^b$.  Note here
\bq\label{01.5}
\int d^4x\sqrt{-g}g^{ad}g^{bc}\na_au_b\na_cu_d=\int d^4x\sqrt{-g}(\na^au^b)(\na^cu^d)
\end{equation}
and also
\bq\label{01.6}
\na_a\left(M^{abcd}\na_c\right)u_d=\na_a\left[g^{ad}g^{bc}-g^{ab}g^{cd}\right]\na_cu_d=
\end{equation}
$$=\na_ag^{ad}g^{bc}\na_cu_d-\na_ag^{ab}g^{cd}\na_cu_d=\na_a\na^bu^a-\na^b\na_cu^c\sim
(\na_a\na^b-\na^b\na_a)u^a$$
Further (as in \eqref{01.6})
\bq\label{01.7}
M^{abcd}\na_au_b\na_cu_d=g^{ad}g^{bc}\na_au_b\na_cu_d-g^{ab}g^{cd}
\na_au_b\na_cu_d=
\end{equation}
$$=\na^du_b\na^bu_d-\na_au^a\na_cu^c$$
which confirms \eqref{01.4}.  We record also from \cite{o2} that
\bq\label{01.8}
(\na_{\mu}\na_{\nu}-\na_{\nu}\na_{\mu})\ga(w)=R(\ga,\pp_{\mu},\pp_{\nu},w)
\end{equation}
which identifies $\na_{\mu}\na_{\nu}-\na_{\nu}\na_{\mu}$ with $R_{\mu\nu}$ and allows us to imagine
\eqref{01.6} as $R_a^bu^a$ with Einstein equations
\bq\label{01.9}
R_a^bu^a=N_a^bu^a\,\,(=N^{bc}g_{ca}g^{ca}u_c)
\end{equation}
for example, which is of course equivalent to $R_{ab}=N_{ab}$ (cf. also \cite{p7}).
Note also $G_{ab}=R_{ab}-(1/2)Rg_{ab}=kT_{ab}$ implies that $R_{\nu}^{\mu}-(1/2)R\gd^{\mu}_{\nu}
=kT_{\nu}^{\mu}$ which upon contraction gives $R=-kT$ (since $\gd^{\mu}_{\mu}=4$) and hence
$R_{ab}=k(T_{ab}-(1/2)Tg_{ab})$.
\\[3mm]\indent
For completeness we sketch here a derivation of the Einstein equations from an action principle
(cf. \cite{a5,c91,m1,w4}).  The Einstein-Hilbert action is $A=\int_{\gO}[{\mf
L}_G+ {\mf L}_M]d^4x$ where ${\mf L}_G=(1/2\chi)\sqrt{-g}{}^4R$ ($\chi=8\pi$ and ${}^4R$ is the
Ricci scalar).  Following \cite{c91} we list a few useful facts first (generally we will
write if necessary $g_{ab}T^{cb}=T^c_{\cdot a}$ and $g_{ab}T^{bc}=T^{\cdot c}_a$).
\begin{enumerate}
\item
$\na_{\gag}g^{\ga\gb}=0$ (by definitions of covariant derivative and Christoffel symbols).
\item
$\gd\sqrt{-g}=(1/2)\sqrt{-g}g^{\ga\gb}\gd g_{\ga\gb}$ and $(\gd g_{\ga\gb})g^{\ga\gb}
=-(\gd g^{\ga\gb})g_{\ga\gb}$ (see e.g. \cite{w4} for the calculation).
\item
For a vector field $v^a$ one has $\na_av^a=\pp_a(\sqrt{-g}v^a)(1/\sqrt{-g})$ and
$\na_{\gb}T^{\ga\gb}=\pp_{\gb}(\sqrt{-g}T^{\ga\gb})(1/\sqrt{-g})+\gG^{\ga}_{\gs\gb}T^{\gs\gb}$
(from $\gG^{\gs}_{\gs\ga}=(1/2)(\pp_{\ga}g_{\mu\nu})g^{\mu\nu}$ and
$\pp_{\ga}(log(\sqrt{-g})=\gG^{\gs}_{\gs\ga}$).
\item
For two metrics $g,\,g^*$ one shows that $\gd\gG^{\ga}_{\gb\gag}=\gG^{*\ga}_{\gb\gag}
-\gG^{\ga}_{\gb\gag}$ is a tensor.
\item
$\gd R_{\ga\gb}=\na_{\gs}(\gd \gG^{\gs}_{\ga\gb}-\na_{\gb}(\gd\gG^{\gs}_{\ga\gs})$
(see \cite{c91} for the calculations).
\item
Recall also Stokes theorem
$\int_{\gO}\na_{\gs}v^{\gs}\sqrt{-g}d^4x=\int_{\gO}\pp_{\gs}(v^{\gs}\sqrt{-g})d^4x=\int_{\pp\gO}
\sqrt{-g}v^{\gs}d^3\gS_{\gs}$.
\end{enumerate}
\indent
Now requiring a stationary action for arbitrary $\gd g^{ab}$ (with certain derivatives of the
$g^{ab}$ fixed on the boundary of $\gO$ one obtains  (${\mf L}_M$ is the matter Lagrangian)
\bq\label{01.10}
\gd
I=\frac{1}{2\chi}\int_{\gO}\left(R_{\ga\gb}-\frac{1}{2}g_{\ga\gb}R\right)\sqrt{-g}\gd
g^{\ga\gb}d^4x+
\end{equation}
$$+\frac{1}{2\chi}\int_{\gO}g^{\ga\gb}\sqrt{-g}\gd R_{\ga\gb}d^4x+\int_{\gO}\frac{\gd{\mf L}_M}
{\gd g^{\ga\gb}}\gd g^{\ga\gb}d^4x=0$$
The second term can be written 
\bq\label{01.11}
\frac{1}{2\chi}\int_{\gO}g^{\ga\gb}\sqrt{-g}\gd R_{\ga\gb}d^4x=\frac{1}{2\chi}\int
g^{\ga\gb}\sqrt{-g}[\na_{\gs}(\gd\gG^{\gs}_{\ga\gb})-\na_{\gb}(\gd\gG^{\gs}_{\ga\gs}]d^4x=
\end{equation}
$$=\frac{1}{2\chi}\int_{\gO}\sqrt{-g}[\na_{\gs}(g^{\ga\gb}\gd\gG^{\gs}_{\ga\gb})-\na_{\gb}
(g^{\ga\gb}\gd\gG^{\gs}_{\ga\gs})]d^4x=$$
$$=\frac{1}{2\chi}\int_{\gO}\pp_{\gs}[(\sqrt{-g}g^{\ga\gb}\gd\gG^{\gs}_{\ga\gb})-(\sqrt{-g}
g^{\ga\gs}\gd\gG^{\rho}_{\ga\rho})]d^4x$$
where $\gd\gG^{\ga}_{\gb\gag}=(1/2)[\na_{\gag}(\gd g_{\gb\gs})+\na_{\gb}(\gd g_{\gs\gag})-
\na_{\gs}(\gd g_{\gag\gb})]$.  This can be transformed into an integral over the boundary
$\pp\gO$ where it vanishes if ceertain derivatives of $g_{\ga\gb}$ are fixed on the boundary.
In fact the integral over the boundary $\pp\gO=\sum S_i$ can be written as $\sum_i(\gep_I/2\chi)
\int_{S_i}\gag_{\ga\gb}\gd \tl{N}^{\ga\gb}d^3x$ where $\gep_i={\bf n}_i\cdot{\bf n}_i=\pm 1$ 
(${\bf n}_i$ normal to $S_i$) and $\gag_{\ga\gb}=g_{\ga\gb}-\gep_i{\bf n}_{\ga}\cdot{\bf
n}_{\gb}$ is the 3-metric on the hypersurface $S_i$ (cf. \cite{y1}).  Further 
\bq\label{01.12}
\tl{N}^{\ga\gb}=\sqrt{|\gag|}(K\gag^{\ga\gb}-K^{\ga\gb})=-\frac{1}{2}g\gag^{\ga\mu}\gag^{\gb\nu}
{\mc L}_{{\bf n}}(g^{-1}\gag_{\mu\nu})
\end{equation}
where $K_{\ga\gb}=-(1/2){\mc L}_{{\bf n}}\gag_{\ga\gb}$ is the extrinsic curvature of each
$S_i$ and ${\mc L}_{{\bf n}}$ is the Lie derivative.  Consequently if the quantities 
$\tl{N}^{\ga\gb}$ are fixed on the boundary for an arbitrary $\gd g_{\ga\gb}$ one gets from
the first and last equations in \eqref{01.10} the Einstein field equations
\bq\label{01.13}
G_{\ga\gb}=R_{\ga\gb}-\frac{1}{2}Rg_{\ga\gb}=\chi T_{\ga\gb};\,\,T_{\ga\gb}=-2\frac{\gd{\mf
L}_M}{\gd g^{ab}}+{\mf L}_Mg_{\ga\gb}
\end{equation}
We note here that 
\bq\label{01.14}
\gd\int {\mf L}_m\sqrt{-g}d^4x=\int\frac{\gd{\mf L}_m}{\gd g^{ab}}\sqrt{-g}d^4x+\int{\mf
L}_m\gd(\sqrt{-g})d^4x=
\end{equation}
$$=\int\frac{\gd{\mf L}_m}{\gd g^{ab}}\sqrt{-g}d^4x-\frac{1}{2}\int{\mf L}_mg_{ab}
(\gd g^{ab})\sqrt{-g}d^4x$$
A factor of 2 then arises from the $2\chi$ in (5.16).
\\[3mm]\indent
{\bf REMARK 5.2.}
Let us rephrase some of this following \cite{w4} for clarity.  Thus e.g. think of
functionals
$F(\psi)$ with $\psi=\psi_{\gl}$ a one parameter family and set $\gd\psi=(d\psi_{\gl}/d\gl)|_
{\gl=0}$.  For $F(\psi)$ one writes then $dF/d\gl=\int \phi\gd\psi$ and sets $\phi=
(\gd F/\gd\psi)|_{\psi_0}$.  Then (assuming all functional derivatives are symmetric with
no loss of generality) one has for ${\mf L}_G=\sqrt{-g}R$ and $S_G=\int {\mf L}_Gd^4x$
\bq\label{01.15}
\frac{d{\mf L}_G}{d\gl}=\sqrt{-g}(\gd R_{ab})g^{ab}+\sqrt{-g}R_{ab}\gd g^{ab}+R\gd(\sqrt{-g})
\end{equation} 
But $g^{ab}\gd R_{ab}=\na^av_a$ for $v_a=\na^b(\gd g_{ab})-g^{cd}\na_a(\gd g_{cd})$.  Further
$\gd\sqrt{-g}=-(1/2)\sqrt{-g}g_{ab}\gd g^{ab}$ so one has
\bq\label{01.16}
\frac{dS_G}{d\gl}=\int\frac{d{\mf L}_G}{d\gl}d^4x=\int\na^av_a\sqrt{-g}d^4x+\int
\left(R_{ab}-\frac{1}{2}Rg_{ab}\right)(\gd g^{ab})\sqrt{-g}d^4x
\end{equation}
Discarding the first term as a boundary integral we get the first term in
(5.16).$\hfill\bs$
\\[3mm]\indent
{\bf REMARK 5.3.}
From \cite{p1} we see that the entropy in S in (5.7) reduces to a 4-divergence when the 
Einstein equations are satisfied ``on shell" making S a surface term
\bq\label{02.77}
S=\frac{1}{8\pi}\int_Vd^4x\sqrt{-g}\na_i(u^b\na_bu^i-u^i\na_bu^b)=
\end{equation}
$$=\frac{1}{8\pi}
\int_{\pp V}d^3x\sqrt{h}n_i(v^b\na_bu^i-u^i\na_bu^b)$$
Thus the entropy of a bulk region V of spacetime resides in its boundary $\pp V$
when the Einstein equations are satisfied.  In varying (5.7) to obtain
(5.8) one keeps the surface contribution to be a constant.  Thus in a
semiclassical limit when the Einstein equations hold to the lowest order
the entropy is contributed only by the boundary term and the system is holographic.
$\hfill\bs$
\\[3mm]\indent
{\bf REMARK 5.4.}
Let us call attention here to \cite{e1,j2} where a very different approach is made
to derive the Einstein equations from thermodynamics using entropy ideas.  Using
non-equilibrium thermodynamics one finds also that entropy dependence on the Ricci
scalar can be accomodated.$\hfill\bs$

\section{WDW AND THE EINSTEIN EQUATIONS}
\renewcommand{\theequation}{6.\arabic{equation}}
\setcounter{equation}{0}

We sketch here the derivation of the Einstein equations from quantum geometrodynamics
following Gerlach \cite{g11}.  He works with the Einstein HJ (EHJ) equation in the 
Perez form (cf. \cite{p11})
\bq\label{6.1}
{}^3R+h^{-1}\left(\frac{1}{2}h_{ij}h_{k\ell}-h_{ik}h_{j\ell}\right)\frac{\gd S}{\gd
h_{ij}}\frac{\gd S}{\gd h_{k\ell}}
\end{equation}
where $h_{ij}$ is the metric of the spatial hypersurface $\gS$.  One defines $({\bf
6A})\,\,\gd S=\int [\gd S/\gd h_{ij}(x)]\gd h_{ij}(x)d^3x$ with integration over 
$\gS$ and assumes that S is a function of the 3-geometry only, namely $({\bf 6B})\,\,
S=S[{}^3{\mf G})$ (i.e. S is coordinate independent).  Assume further the principle
of constructive interference (see below) and that either $\gS$ is finite with no boundary
or that $\gS$ is asymptotically flat.  Under these conditions one proves that there
are 4 functions $N,\,N_i\,\,(i=1,2,3)$ which together with $h_{ij}$ give a spacetime
metric
\bq\label{6.2}
ds^2=h_{ij}(N^idx^0+dx^i)(N^jdx^0+dx^j)-N^2(dx^0)^2=
\end{equation}
$$=h_{ij}dx^idx^j+2N_idx^idx^0+
(N_jN^j-N^2)(dx^0)^2$$
which satisfies the Einstein field equations.  Further the manifestly covariant 
equations of geometrodynamics 
\bq\label{6.3}
\frac{\gd h_{ij}}{\gd\gs}=\frac{\gd H}{\gd\pi^{ij}(x)};\,\,\frac{\gd \pi^{ij}(x)}{\gd\gs}
=-\frac{\gd H}{\gd h_{ij}(x)}
\end{equation}
hold where $H=H[h_{ij}]$ and $\pi^{ij}=(\gd S/\gd h_{ij})$.  Here $\gs$ is the 
Tomonaga-Schwinger many fingered time parameter (cf. \cite{c1,n2,s2,t1}).
One notes that the dynamical phase $S=S[h_{ij}]$ is required to be a functional of the
3-geometry alone, regardless of coordinates so one writes $({\bf 6C})\,\,
S=S[{}^3{\mf G}]$ which means that $({\bf 6D})\,\,\na_j[\gd S/\gd h_{ij}]=0$.
To see this consider $h_{ij}(x)$ with $x^i\to x^{'i}=x^i+\gep \xi^i(x)$ while 
preserving the geometry where 
\bq\label{6.4}
h'_{ij}(x)=h_{ij}(x)+\gd h_{ij}(x);\,\,\gd h_{ij}(x)
=-\gep(\na_j\xi_i+\na_i\xi_j)
\end{equation}
The ostensible change in S would be
\bq\label{6.5}
\gd S=\int \frac{\gd S}{\gd h_{ij}(x)}\gd h_{ij}(x)d^3x=
\end{equation}
$$=-2\gep\int\frac{\gd S}{\gd
h_{ij}(x)}\na_j\xi_id^3x=2\gep\int \na_j\left(\frac{\gd S}{\gd
h_{ij}(x)}\right)\xi_id^3x$$
However the ${}^3{\mf G}$ itself is not changed so $\gd S$ must vanish which means that
$({\bf 6D})$ holds.
\\[3mm]\indent
Now the phase functional is defined on superspace (i.e. the set of equivalence classes
of spacelike $h_{ij}(x)$ that can be transformed into each other by spatial coordinate
transformations.  One considers a solution to the EHJ equation satisfying
({\bf 6D}), namely $S[{}^3{\mf G};\ga(u),\gb(u)]$ where $\ga,\,\gb$ are integration
constants defined by parameters $u=(u_1,u_2,u_3)$.  Then consider the phase functionals
obtained via $({\bf 6E})\,\,\ga(u)\to \ga(u)+\gd\ga(u)$ and $\gb(u)\to \gb(u)+\gd\gb
(u)$.  Some argument (cf. \cite{g11}) yields then $({\bf 6F})\,\,(\gd S/\gd{}^3{\mf G})
(\ga+\gd\ga,\gb+\gd\gb)d{}^3{\mf G}=(\gd S/\gd{}^3{\mf G})(\ga,\gb)d{}^3{\mf G}$.
One is concerned here not with e.g. $\gd h_{ij}/\gd\gs$ where
$\gs$ is the Tomonaga time parameter (discussed in Section 6.1 below) but rather
with a parametrization independent quantity such as
$({\bf 6G})\,\,\gd{}^2{\mf G}=\int (\gd{}^3{\mf G}/\gd\gs)\gd\gs d^3x$ or
equivalently with $\gd h_{ij}=\int (\gd h_{ij}/\gd\gs)\gd\gs d^3x$.  Thus the focus
of attention is
$\gd h_{ij}$ rather than $\gd h_{ij}/\gd \gs$ and this allows one to forego the details
of some messy parametrization scheme.  Now from ({\bf 6F}) a necessary condition for
the  ``vector" $\gd{}^3{\mf G}/\gd\gs(x)$ to be tangent to a history through
${}^3{\mf G}$ is $({\bf 6H})\,\,\gd(\gd S/\gd{}^3{\mf G})(\gd{}^3{\mf G}/\gd \gs)=0$
where
$(\gd S/\gd{}^3{\mf G})$ denotes the change due to an arbitrary infinitesimal
variation in $(\ga(u),\gb(u))$.  Then the EHJ equation together with ({\bf 6H})
contains all of  general relativity (GR), exhibited then as (see below)
\bq\label{6.6}
{}^3R-\left(\frac{\gd S}{\gd{}^3{\mf G}}\right)\left(\frac{\gd S}{\gd{}^3{\mf G}}\right)^*
=0;\,\,\gd\left(\frac{\gd S}{\gd{}^3{\mf G}}\right)\frac{\gd{}^3{\mf G}}{\gd\gs}=0
\end{equation}
The starred vector $(\gd S/\gd{}^3{\mf G})^*$ is the dual with respect to the deWitt
metric and an easy way now of obtaining the Einstein field equations (Efe) is to use the
language of tensor analysis.  Then the tangent vector in ({\bf 6H}) becomes
\bq\label{6.7}
\frac{\gd{}^3{\mf G}}{\gd\gs(x')}\to\frac{\gd
h_{ij}(x,\gs(x'))}{\gd\gs(x')};\,\,\frac{\gd S}{\gd{}^3{\mf G}}\to\frac{\gd S}{\gd
h_{ij}(x)}\equiv\pi^{ij}(x)
\end{equation}
where $({\bf 6I})\,\,\na_j\pi^{ij}=0$ in order for S to depend only on ${}^3{\mf G}$.
With this notation \eqref{6.6} becomes
\bq\label{6.8}
\int \gd\pi^{ij}(x)\frac{\gd h_{ij}(x)}{\gd\gs(x')}d^3x=0;\,\,{}^3R+h^{-1}
\left(\frac{1}{2}h_{ij}h_{k\ell}-h_{ik}h_{j\ell}\right)\pi^{ij}\pi^{k\ell}=0
\end{equation}
One enunciates the principle of constructive interference stated in \eqref{6.8}
now as follows:  In order that a change $\gd h_{ij}$ or equivalently 
$\gd h_{ij}/\gd\gs$ be a vector tangent to a history it is necessary that there 
exist a $\pi^{ij}(x)$ with the property
\bq\label{6.9}
\int\pi^{ij}(x)\frac{\gd h_{ij}(x)}{\gd\gs (x')}d^3x=extremum
\end{equation}
if one changs the integration constants $\ga(u),\,\gb(u)$ slightly.  One discusses
the fact that freedom in adjusting the integration constants corresponds to freedom
in choosing the momentum density and relates this to the idea of having a complete
solution S as a functional of the maximum number of possible independent 
constants (cf. \cite{g11} for details).
\\[3mm]\indent
Now going to the Efe one replaces \eqref{6.9} with the help of ({\bf 6G}) by writing
$({\bf 6J})\,\,\int \pi^{ij}\gd h_{ij}d^3x$ and this must be an extremum with
respect to variations in $\pi^{ij}(x)$ subject to the restrictions
\bq\label{6.10}
R_0=h^{1/2}\left[{}^3R+h^{-1}\left(\frac{1}{2}h_{ij}h_{k\ell}-h_{ik}h_{j\ell}\right)
\pi^{ij}\pi^{k\ell}\right]=0;
\end{equation}
$$\na_j\pi^{ij}=0;\,\,\pi^{ij}(x)=\frac{\gd S}{\gd h_{ij}}$$
One can take the restrictions on S into account in the extremum principle by
multiplying them by yet to be determined functions $\gd M(x)$ and $2\gd M_i(x)$
and add to ({\bf 6J}) to get $\int [\pi^{ij}\gd h_{ij}+\gd M R_0+2\gd
M_i\na_j\pi^{ij}] d^3x$.  Now consider changes in the integral due to arbitrary
variations in
$\pi^{ij}$; an integration by parts yields
\bq\label{6.11}
\int_{\gS} \left[\gd h_{ij}\gd\pi^{ij}+\gd M\left(\frac{\gd R_0}{\gd
\pi^{ij}}\right)\gd\pi^{ij}-2\gd \na_jM_i\gd\pi^{ij}\right]d^3x+ 
\int_{\pp\gS}\gd M_i\gd\pi^{ij}dS_j
\end{equation}
The surface term vanishes due to boundary conditions and one emphasizes that the
change $\gd h_{ij}$ has nothing to do with the variations in $\pi^{ij}$.  The
arbitrary changes in $\pi^{ij}$ fall into two classes ({\bf A}) Those that satisfy
the variation equations ({\bf 6I}) and \eqref{6.8} and ({\bf B}) Those that do not.
The principle of constructive interference requires that the variations of the
integral \eqref{6.11} vanish for class ({\bf A}) variations.  Consequently
the coefficients of these variations must vanish and one then adjusts the functions
$\gd M$ and $\gd M_i$ so that
the coefficients of the class ({\bf B}) variations also vanish.  The result of this
is
\bq\label{6.12}
\gd h_{ij}=-2\gd M(h)^{-1/2}\left(\frac{1}{2}g_{ij}\pi^k_k-\pi^{ij}\right)+
(\gd \na_jM_i+\na_iM_j)
\end{equation}
This equation relates the change in $h_{ij}$ between two close 3-geometries to the 
momentum $\pi^{ij}$ and is discussed in \cite{g11}.  In order to put this in a more
familiar form one notes that
\bq\label{6.13}
\gd M=\int\frac{\gd M}{\gd\gs(x')}\gd\gs(x')d^3x';\,\,\gd M_i=\int\frac{\gd M_i}
{\gd\gs(x')}\gd\gs(x')d^3x'
\end{equation}
Hence
\bq\label{6.14}
\frac{\gd h_{ij}}{\gd\gs(x')}=-2\frac{\gd
M(x,\gs)}{\gd\gs(x')}g^{-1/2}\left(\frac{1}{2}g_{ij}\pi^k_k-\pi_{ij}\right)+
\frac{(\gd\na_jM_i+\gd\na_iM_j)}{\gd\gs(x')}
\end{equation}
which amounts to
\bq\label{6.15}
\frac{\gd M(x,\gs)}{\gd\gs(x')}[(1/2)g_{ij}\pi^k_k-\pi_{ij})=\frac{1}{2}
\left(\frac{\gd \na_jM_i}{\gd\gs(x')}+\frac{\gd\na_iM_j}{\gd\gs(x')}-
\frac{\gd h_{ij}}{\gd\gs(x')}\right)g^{1/2}
\end{equation}
These equations are still manifestly covariant and by introducing
\bq\label{6.16}
H_0(x')=-\int\frac{\gd M(x)}{\gd \gs(x')}R_0(x)d^3x;\,\,H_1(x')=-2
\int\frac{\gd M_i(x)}{\gd \gs(x')}\na_j\pi^{ij}(x)d^3x
\end{equation}
one can rewrite \eqref{6.14} as
\bq\label{6.17}
\frac{\gd h_{ij}(x,\gs)}{\gd\gs(x')}=\frac{\gd(H_0(x')+H_1(x'))}{\gd \pi^{ij}(x)}
\end{equation}
Note that $h_{ij}(x,\gs)$ is a functional of $\gs(x')$ and introduce a particular
parameter for the hypersurface, say $\gs(x')=t$, in which case
\bq\label{6.18}
\frac{\pp h_{ij}(x,,t)}{\pp t}=\int\frac{\gd h_{ij}(x,\gs)}{\gd\gs(x')}d^3x'
\end{equation}
Integrating \eqref{6.14}-\eqref{6.15} in $x'$ gives
\bq\label{6.19}
\frac{\pp h_{ij}}{\pp t}=-2N(h)^{-1/2}[(1/2)g_{ij}\pi^k_k-\pi_{ij})+2\na_jN_i
\equiv 
\end{equation}
$$\equiv (1/2)h_{ij}\pi^k_k-\pi_{ij}=h^{1/2}(\na_jN_i+\na_iN_j-\pp_th_{ij})/2N$$
where
\bq\label{6.20}
\na_jN_i=\int\frac{\gd \na_jM_i}{\gd\gs(x')}d^3x';\,\,N=\int\frac{\gd
M}{\gd\gs(x')}d^3x'
\end{equation}
Now two conclusions can be drawn from \eqref{6.12} and \eqref{6.14}
\begin{itemize}
\item
The term $\na_jN_i$ transforms like a 3-tensor so $N_i$ is a covariant 3-vector
\item
The factor N transforms like a 3-scalar.
\end{itemize}
In addition the two equations serve two purposes
\begin{itemize}
\item
\eqref{6.12} reveals how a tangent vector $\gd{}^3{\mf G}/\gd\gs(x')$ must be
related to $\pi^{ij}=\gd S/\gd h_{ij}$ if this vector is tangent to its history
\item
\eqref{6.19} serves as a definition of the extrinsic curvature if one sets
\bq\label{6.21}
h^{-1/2}[(1/2)g_{ij}\pi^k_k-\pi_{ij})]=K_{ij}
\end{equation}
provided that one identifies the hypersurface parameter t with the fourth coordinate
and the functions N and $N_i$ with the lapse and shift functions (cf. \eqref{6.2})
\end{itemize}
Having determined how $h_{ij}$ varies along a classical history (half of the
dynamical equations) one does the same thing for $\pi^{ij}$ via
\bq\label{6.22}
\frac{\gd\pi^{ij}}{\gd\gs}=\int\frac{\gd\pi^{k\ell}(x)}{\gd h_{ij}(x')}\frac{\gd
h_{ij}(x')} {\gd\gs}d^3x'=\int \frac{\gd \pi^{ij}(x')}{\gd h_{k\ell}(x)}\frac
{\gd h_{ij}(x')}{\gd\gs}d^3x
\end{equation}
The EHJ equation \eqref{6.1} holds for all ${}^3{\mf G}$ and hence the functional
derivatives of \eqref{6.1} and ({\bf 6D}) with respect to $h_{ij}(x)$ must vanish
at all functions $h_{ij}$, so
\bq\label{6.23}
0=\int\frac{\gd H_0}{\gd\pi^{ij}(x')}\frac{\gd \pi^{ij}(x')}{\gd h_{k\ell}(x)}d^3x'
+\frac{\gd H_0}{\gd h_{k\ell}(x)};
\end{equation}
$$0=\int\frac{\gd H_1}{\gd\pi^{ij}(x')}\frac{\gd\pi^{ij}(x')}{\gd
h_{k\ell}(x)}d^3x'+\frac{\gd H_1}{\gd h_{k\ell}(x)}$$
To evaluate the expression on the right in \eqref{6.22} put in \eqref{6.17} for
$\gd h_{ij}/\gd\gs$ to get
\bq\label{6.24}
\frac{\gd\pi^{k\ell}(x)}{\gd\gs}=\int\frac{\gd\pi^{ij}(x')}{\gd h_{k\ell}(x)}\left(
\frac{\gd H_0}{\gd\pi^{ij}(x')}+\frac{\gd H_1}{\gd\pi^{ij}(x')}\right)d^3x'
\end{equation}
But via \eqref{6.22} the right side of this reduces to
\bq\label{6.25}
\frac{\gd\pi^{k\ell}(x)}{\gd\gs}=-\frac{\gd(H_0+H_1}{\gd h_{k\ell}}
\end{equation}
Hence the change in $\pi^{k\ell}$ for a given test function $\gd\gs(x')$ is
\bq\label{6.26}
\gd\pi^{k\ell}(x)=-\int\frac{\gd[H_0(x')+H_1(x')]}{\gd h_{k\ell}(x)}\gd \gs(x')d^3x'
\end{equation}
The ensuing momentum equations \eqref{6.25}-\eqref{6.26} are also manifestly
covariant.
\\[3mm]\indent
One has now obtained 3 constraint equations ({\bf 6D}) and two sets of equations
\eqref{6.17} and \eqref{6.25}; it remains to show that these plus the EHJ equation
are equivalent to the ten Efe.  First it is shown (cf. \cite{g11}) that
$\gd\pi^{ij}+\pi^{ij}$ is a legitimate momentum in that it satisfies \eqref{6.1}
and ({\bf 6D}).  Then write the available equations (\eqref{6.8},
\eqref{6.17}, \eqref{6.25}, and ({\bf 6I}) together as
\bq\label{6.27}
{}^3R+h^{-1}[(1/2)h_{ij}h_{k\ell}-h_{ik}h_{j\ell}]\pi^{ij}\pi^{k\ell}=0;\,\,
\na_j\pi^{ij}=0;
\end{equation}
$$\frac{\gd h_{ij}(x)}{\gd\gs(x')}=\frac{\gd[H_0(x')+H_1(x')]}{\gd\pi^{ij}(x)};
\,\,\frac{\gd\pi^{ij}}{\gd\gs(x')}=-\frac{\gd[H_0(x')+H_1(x')]}{\gd h_{ij}(x)}$$
However the first 2 equations are essentially contained in the last 2 equations
(once this holds at the initial point).  The last two equations are covariant and
hold on every 3-D slice through spacetime.  That the above four equations imply the
ten Efe can be best seen by observing that these equations can be derived from a
variational principle whose Lagrangian is
\bq\label{6.28}
{\mf L}=\int\left[\frac{\gd h_{ij}}{\gd\gs(x')}\pi^{ij}+\frac{\gd
M}{\gd\gs(x')}\times
[h^{1/2}\,{}^3R+h^{-1/2}[(1/2)\pi_i^i\pi_j^j-\pi_{ij}\pi^{ij})]+\right.
\end{equation}
$$\left.+2\frac{\gd M_i}{\gd\gs(x')}\na_j\pi^{ij}-2\left(\pi^{ij}\frac{\gd M_j}
{\gd\gs(x')}-\frac{1}{2}\pi_k^k\frac{\gd M^i}{\gd\gs(x')}+h^{1/2}\frac
{\gd M^{,i}}{\gd\gs(x')}\right)_{,i}\right]d^3x'=$$
$$=\pp_th_{ij}\pi^{ij}+N[h^{1/2}\,{}^3R+h^{-1/2}((1/2)\pi_i^i\pi^j_j-\pi_{ij}\pi^{ij})]
-2N_i\na_j\pi^{ij}-$$
$$-2(\pi^{ij}N_j-(1/2)\pi_k^kN^i+h^{1/2}N^{,i)})_{,i}$$
(the notation $f_{,i}$ presumably means $\pp_if$ ?).
This Lagrangian for the 3+1 formulation is equal to $({\bf 6K})\,\,{\mf
L}=(-{}^4g)^{1/2}\,{}^4R$.  The necessary identifications with the 4-geometry are then
\bq\label{6.29}
h_{ij}={}^4g_{ij};\,\,N=(-{}^4g^{00})^{1/2};\,\,N_i={}^4g_{0i};
\end{equation}
$$\pi^{ij}=h({}^4\gG^0_{mn}-h_{mn}{}^4\gG_{k\ell}h^{k\ell}){}^4g^{im}\,{}^4g^{jn};
\,\,(Nh)^{1/2}=(-{}^4g)^{1/2}$$
Denoting the Efe by $G_{\mu\nu}=0\,\,(\mu,\nu=0,1,2,3)$ then \eqref{6.8} and 
({\bf 6I}) are $G^0_{\nu}=0$ while \eqref{6.25} is a linear combination of these
equations together with the remaining 6 Efe where \eqref{6.17} serves as the
definition of $\pi^{ij}(x)$.  Putting in now $\psi=exp(iS/\hbar)$ in superspace,
S is the solution of the EHJ equation ${}^3R-(\gd S/\gd{}^3{\mf G})(\gd
S/\gd{}^3{\mf G})^*=0$.

\subsection{MULTIFINGERED TIME}

The discussion of the multifingered time (MFT) of Tomonaga in \cite{g11} can be
improved as in \cite{n2} (cf. also \cite{c1,s2,t1}).  Let $x=\{x^{\mu}\}=(x^0,{\bf
x})$ be spacetime coordinates.  A timelike Cauchy hypersurface $\gS$ can be defined
via a function $T(x)$ via the equation ({\bf (6M}) $x^0=T({\bf x})$.  If $T({\bf
x})$ is given then ${\bf x}\in\gS$ is correct and if $\gs\subset\gS$ then
e.g. $T_{\gs}$ denotes the set of values for ${\bf x}\in\gs$.  For a scalar field
$\phi$ one describes its dynamics via 
\bq\label{6.31}
\hat{H}({\bf x})\psi[\phi,T]=i\frac{\gd\psi[\phi,T]}{\gd T({\bf x})}
\end{equation}
A wave functional $\psi[\phi,T]$ can be viewed as a functional of $\phi_{\gS}$
and \eqref{6.31} shows how $\psi$ changes for an infinitesimal change $\gd T(x)$
(we will ocassionally omit boldface on $x$ now).  Thus \eqref{6.31} is a generalized
SE but it does not involve any preferred foliation of spacetime.  Since $\gS$ is
determined by T one can say that $\rho[\phi,T]=|\psi[\phi,T]|^2$ is the probability
density for the field to have the value $\phi$ at time T but remember that T is a
collection of real parameters with one real parameter for each point ${\bf x}$.
Consider now a free scalar field with Hamiltonian density
$({\bf
6N})\,\,\hat{H}(x)=-(1/2)(\gd^2/\gd\phi^2(x))+(1/2)[(\na\phi(x))^2+m^2\phi^2(x)]$.
Then writing $\psi=Rexp(iS)$ one obtains
\bq\label{6.32}
\frac{1}{2}\left(\frac{\gd
S}{\gd\phi(x)}\right)^2+\frac{1}{2}[(\na\phi(x))^2+m^2\phi^2(x)]+{\mf Q}(x,\phi,T)
+\frac{\gd S}{\gd T(x)}=0
\end{equation}
\bq\label{6.33}
\frac{\gd\rho}{\gd T(x)}+\frac{\gd}{\gd\phi(x)}\left(\rho\frac{\gd
S}{\gd\phi(x)}\right)=0;\,\,{\mf Q}=-\frac{1}{2R}\frac{\gd^2R}{\gd \phi^2(x)}
\end{equation}
The Bohmian interpretation involves a deterministic time dependent hidden variable
such that the time evolution of this variable is consistent with the probabilistic
interpretation of $\rho$.  This is naturally achieved by introducing a MFT field
$\Phi(x,T)$ satisfying the MFT Bohmian equation of motion
\bq\label{6.34}
\frac{\gd\Phi(x,T)}{\gd T(x')}=\gd^3(x-x')\left.\frac{\gd S}{\gd\phi(x)}\right|_{\phi
=\Phi};\,\,\int_{\gs_x}d^3x'\frac{\gd\Phi(x,T)}{\gd T(x')}=\left.\frac{\gd S}
{\gd \phi(x)}\right|_{\Phi=\phi}
\end{equation}
where $\gs_x$ is an arbitrarily small region around $x$.  The second
equation in \eqref{6.34} is the MFT version of the usual single-time Bohmian equation
of motion
$\pp_t\Phi(x,t)= (\gd S/\gd\phi(x))|_{\phi=\Phi}$ whereas the first equation is
more fundamental since no $\gs_x$ is involved.  For comparison purposes however
integration within $\gs_x$ is useful; e.g. using \eqref{6.32} and \eqref{6.34}
one has
\bq\label{6.35}
\left[\left(\int_{\gs_x}d^3x'\frac{\gd}{\gd
T(x')}\right)^2-\na^2_x+m^2\right]\Phi(x,T)=-\left.\int_{\gs_x}d^3x'\frac{\gd{\mf
Q}(x',
\phi,T)}{\gd\phi(x)}\right|_{\phi=\Phi}
\end{equation}
This can be viewed as an MFT Klein-Gordon equation with a quantum term added.
Note that officially one should write $\Phi({\bf x},T({\bf x}))=\phi({\bf x},x^0)=
\Phi(x)$ and we assume this is understood throughout.  
\\[3mm]\indent
Now to provide a manifestly covariant QFT one introduces ${\bf s}=(s^1,s^2,s^3)$
which serve as coordinates on a 3-D manifold; then write $x^{\mu}=X^{\mu}({\bf s})$
leading to one equation $f(x^0,x^1,x^2,x^3)=0$ determining a 3-D hypersurface in
spacetime.  Assume a background metric $g_{\mu\nu}(x)$ is given with induced metric
\bq\label{6.36}
h_{ij}(s)=g_{\mu\nu}(X(s))\frac{\pp X^{\mu}(s)}{\pp s^i}\frac{\pp X^{\nu}(s)}{\pp
s^j}
\end{equation}
on the hypersurface.  A normal and unit normal to this surface is then
\bq\label{6.37}
\tl{n}_{\mu}(s)=\gep_{\mu\ga\gb\gag}\frac{\pp X^{\ga}}{\pp s^1}\frac{\pp X^{\gb}}
{\pp s^2}\frac{\pp X^{\gag}}{\pp s^3};\,\,n^{\mu}(s)=
\frac{g^{\mu\nu}\tl{n}_{\nu}}{\sqrt{|g^{\ga\gb}\tl{n}_{\ga}\tl{n}_{\gb}}}
\end{equation}
Now the equations above can be written in a covariant form via
\bq\label{6.38}
{\bf x}\to{\bf s};\,\,\frac{\gd}{\gd T({\bf x})}\to\frac{\gd}{\gd\tau({\bf s})}
\equiv n^{\mu}({\bf s})\frac{\gd}{\gd X^{\mu}({\bf s})}
\end{equation}
The Tomonaga-Schwinger equation \eqref{6.31} becomes then
\bq\label{6.39}
\hat{H}({\bf s})\psi[\phi,X]=in^{\mu}({\bf s})\frac{\gd\psi[\phi,X]}{\gd
X^{\mu}({\bf s})}
\end{equation}
and for free fields the Hamiltonian density operator in curved spacetime is
\bq\label{6.40}
\hat{H}=\frac{-1}{2|h|^{1/2}}\frac{\gd^2}{\gd\phi^2({\bf s})}+\frac{|h|^{1/2}}
{2}[-h^{ij}(\pp_i\phi)(\pp_j\phi)+m^2\phi^2]
\end{equation}
The Bohmian equations of motion \eqref{6.34} become
\bq\label{6.41}
\frac{\gd\Phi(s,X)}{\gd\tau({\bf s}'}=\left.\frac{\gd^3({\bf s}-{\bf s}')}{|h({\bf
s})|^{1/2}}\frac{\gd S}{\gd\phi({\bf s})}\right|_{\phi=\Phi}
\end{equation}
and \eqref{6.35} becomes
\bq\label{6.42}
\left[\left(\int_{\gs_x}d^3s'\frac{\gd}{\gd\tau(s')}\right)^2+\na^i\na_i
+m^2\right]\Phi(s,X)=-\left.\int_{\gs_x}\frac{d^3s'}{\sqrt{|h|}}\frac{\gd{\mf Q}
(s',\phi,X)}{\gd\phi(s)}\right|_{\phi=\Phi}
\end{equation}
where $\na_i$ is the covariant derivative in $s^i$ and 
\bq\label{6.43}
{\mf Q}({\bf s},\phi,X)=-\frac{1}{\sqrt{|h({\bf s})|}}\frac{1}{2R}\frac{\gd^2R}
{\gd\phi^2({\bf s})}
\end{equation}
There is a sort of gauge freedom associated related to the covariance due to the 
freedom in choosing the $X^{\mu}({\bf s})$.  For a timelike hypersurface the simplest
choice of gauge is $X^i({\bf s})=s^i$.  This choice implies $\gd X^i({\bf s})=0$
which leads to some of the previous equations prior to covariance.  For example
\eqref{6.41} becomes
\bq\label{6.44}
(g^{00}({\bf x}))^{1/2}\frac{\gd\Phi(x,X^0)}{\gd X^0(x')}=\left.\frac{\gd^3({\bf
x}-{\bf x}')}{|h({\bf x})|^{1/2}}\frac{\gd S}{\gd\phi({\bf x})}\right|_{\phi=\Phi}
\end{equation}
which is the curved spacetime version of \eqref{6.34}.
The covariant formulation of QFT leads to a covariant MFT Bohmian interpretation of
quantum fields which also does not involve a preferred foliation of spacetime.
The covariant Bohmian dynamics does not depend on the choice of coordinates but when
a choice is made then the solution of the MFT Bohmian equations of motion can be
written so that the MFT nature of the field is not manifest.  However the Bohmian
equation of motion retains its covariant form.

\section{TIME}
\renewcommand{\theequation}{7.\arabic{equation}}
\setcounter{equation}{0}

We have seen how MFT arises in QFT and we want to examine this further in connection
with gravity.  We begin with remarks based on
\cite{a11,a5,b4,b5,b3,d2,e1,g5,g4,g3,k4,k5,k2,k37,n1,n2,n3,p85,s23,w2,w1,y1,y2}.  We note
first that time can arise naturally for WDW when using the dDW theory but a MFT approach
seems to require the semiclassical approach and some interaction with matter (see however
\cite{g11} as discussed in Section 6).  Weakening the
Hamiltonian constraint as in \cite{n3} (discussed in Remark 3.1) also provides a time.
The semiclassical approach is illustrated in \cite{b5,g11,k2} for example and we will
sketch some of this here following \cite{g3}.
\\[3mm]\indent
We forgoe the sandwich ordering here for convenience - it remains our principal
ordering candidate however.  Thus consider ($c=1$)
\bq\label{7.1}
H\psi[h_{ab},\phi]=\left(-16\pi G\hbar^2G_{abcd}\frac{\gd^2}{\gd h_{ab}\gd h_{cd}}-
\frac{\sqrt{h}}{16\pi G}(R-2\gL)+H_m\right)\psi=0
\end{equation}
The integrated form of \eqref{7.1} is
\bq\label{7.2}
\int d^3xNH\psi\equiv H^N\psi=(H_G^N+H_m^N)\psi=0
\end{equation}
One uses now an Ansatz ($M=32\pi G)^{-1}$)
\bq\label{7.3}
\psi=exp\left[i(MS_0+S_1+M^{-1}S_2+\cdots)/\hbar\right]
\end{equation}
leading to a set of equations of consecutive orders in M.  The highest order $M^2$
shows that $S_0$ depends only on the 3-metric h (cf. \cite{k2}) and the next order M
gives the HJ equation for the gravitational field
\bq\label{7.4}
H_x=\frac{1}{2}G_{abcd}\frac{\gd S_0}{\gd h_{ab}}\frac{\gd S_0}{\gd h_{cd}}-
2\sqrt{h}(R-2\gL)=0
\end{equation}
Note that these depend on the lapse function $N(x)$.  At the next order $M^0$ it is
convenient to introduce a functional $({\bf 7A})\,\,\psi=D(h_{ab})exp(iS_1/\hbar)$ and
require that D satisfies
\bq\label{7.5}
G_{abcd}\frac{\gd S_0}{\gd h_{ab}}\frac{\gd D}{\gd h_{cd}}-\frac{1}{2}G_{abcd}
\frac{\gd^2S_0}{\gd h_{ab}\gd h_{cd}}D=0
\end{equation}
(note D corresponds to the vanVleck determinant).
The important observation here is that $\psi$ obeys the equation
\bq\label{7.6}
i\hbar G_{abcd}\frac{\gd S_0}{\gd h_{ab}}\frac{\gd\psi}{\gd h_{cd}}=H_m\psi
\end{equation}
which can be rewritten in terms of vector fields
\bq\label{7.7}
\chi(x)=G_{abcd}\frac{\gd S_0}{\gd h_{ab}(x)}\frac{\gd}{\gd h_{cd}}=-2K_{cd}\frac
{\gd}{\gd h_{cd}(x)};\,\,K_{cd}=-\frac{1}{2}G_{abcd}\frac{\gd S_0}{\gd h_{ab}}
\end{equation}
where $K_{cd}$ has the meaning of an extrinsic curvature.  If one now writes
$({\bf 7B})\,\,\chi(x)=(\gd/\gd\tau(x))$ then \eqref{7.6} would be a Tomonaga-Schwinger
equation with respect to the MFT $\tau(x)$ (note $\tau$ is really a function on
$Riem(\gS)$).  However this leads to a contradiction since $[(\gd/\gd\tau(x),\gd/
\gd\tau(y)]=0$ of necessity but $[H_m(x),H_m(y)]\ne 0$.  One writes then
$({\bf 7C})\,\,i\hbar\chi^N=H_m^N\psi$ and (with some argument) shows that
in fact
\bq\label{7.14}
[\chi^N,\chi^M]=-2\int_x(N\pp_aM-M\pp_aN)\na_b\left(\frac{\gd}{\gd h_{ab}}\right)=
\int {\mf L}_Kh_{ab}\frac{\gd}{\gd h_{ab}}
\end{equation}
where $({\bf 7D)})\,\,K^a=h^{ab}(N\pp_bM-M\pp_bN)$.  Hence $[\chi^N,\chi^M]\ne 0$ and
time functions as above can never be introduced (because the Ricci scalar R is not
ultralocal in $h_{ab}$).  The vector fields $\chi^N$ are generators of a hypersurface
deformation normal to itself and the commutator generates stretchings of the
hypersurface.
A proper understanding of \eqref{7.14} and its compatibility with \eqref{7.6} 
is obtained however if one expands the diffeomorphism constraints in powers of 
G (or M) which gives $({\bf 7E})\,\,2h_{bc}D_a(\gd S_0/\gd h_{ab})=0$
(cf. \cite{k1} for notation - $D_a\sim$ covariant derivative).  The highest order M yields
(since
$S_0$ does not depend on the scalar field $\phi$) $({\bf 7F})\,\,2h_{bc}D_a(\gd S_0/\gd
h_{ab})=0$ (diffeomorphism invariance of $S_0$).  The next order $M^0$ leads to a
condition on 
$\psi$, namely
\bq\label{7.15}
2h_{bc}D_a\left(\frac{\gd\psi}{\gd h_{ab}}-\frac{\psi}{D}\frac{\gd D}{\gd
h_{ab}}\right)=\phi_{,c}\frac{\gd\psi}{\gd \phi}
\end{equation}
Since D depends only on the 3-metric (cf. \eqref{7.5}) it is appropriate to demand that
it be diffeomorphism invariant by itself, i.e. $({\bf 7G})\,\,h_{bc}D_a(\gd D/\gd
h_{ab})=0$.  From \eqref{7.15} one finds then $({\bf 7H})\,\,2h_{bc}D_a(\gd \psi/\gd
h_{ab})=\phi_{,x}(\gd\psi/\gd\phi)$ which is of the same form as the general solution
$({\bf 7E})$.  Thus it expresses the invariance of the wave functional
$\psi[h_{ab},\phi]$ with respect to simultaneous diffeomorphisms of the metric and
matter field.  The consistency condition condition for ({\bf 7C}) is $({\bf 7I})\,\,
[\chi^N,\chi^M]\psi=[H_m^M,H_m^N]\psi$.
This however is nothing but the momentum constraint in this order of approximation,
namely ({\bf 7H}), since $[\chi^N,\chi^M]$ generates a diffeomorphism of the metric,
(7.9), and $[H_m^M,H_m^N]$ closes on the momentum density of matter which
generates a diffeomorphism of the matter field.  Thus in the full theory the momentum
constraints provide the integrability conditions for the Tomonaga-Schwinger equations
({\bf 7C}).
\\[3mm]\indent
In the explicit case of a scalar field one has e.g.
\bq\label{7.16}
[H_m^M,H_m^N]=-\int_x(N\pp_aM-M\pp_aN)h^{ab}\phi_{,b}\frac{\gd}{\gd\phi}
\end{equation}
Although a family of time functions $\tau(x)$ on $Riem(\gS)$ does not exist one can
integrate ({\bf 7C}) along the vector field $\chi^N$ for one particular choice of N
and this defines a global time parameter t with respect to which one global SE
can be written down.  It is in this sense that QFT with respect to a chosen foliation
emerges from full quantum gravity.  If there are no such general time functions on
$Riem(\gS)$ what about $S(\gS)=Riem(\gS)/Diff(\gS)$?  To answer this one projects the
vector fields $\chi^N$ to S which is possible since $\chi^N$ is invariant under
diffeomorphisms - referring to \cite{g3} for details one arrives at
\bq\label{7.17}
\pi_*[\chi^N,\chi^M]=[\pi_*\chi^N,\pi_*\chi^M]=0
\end{equation}
and there exist functions $\bar{\tau}^N$ on S such that $({\bf 7J})\,\,
\bar{\chi}^N=\gd/\gd\bar{\tau}^N$ where $\bar{\chi}=\pi_*\chi$, etc.  However the
WDW operator is only defined on $Riem(\gS)$ and some of the intervening calculations
do not make sense on S.  There is further discussion of anomalies, etc. that is worth
reading.  This paper corrects some confusion about the existence of Tomonaga-Schwinger
times on $Riem(\gS)$ in other papers (e.g. \cite{b5,k2}) and one should also
exercise caution in this respect relative to the calculations from \cite{g11}
in Section 6.

\subsection{EXTRINSIC CURVATURE AND TIME}

We go now to some papers \cite{a11,a5,g5,g4,k4,p17,y1,y2} where from \cite{a11} 
one recalls that it is not N but the slicing density $\ga(x,t)=Nh^{-1/2}$ is the freely
specifiable quantity for the lapse.  One writes then
\bq\label{7.12}
ds^2=-N^2dt^2+h_{ij}(dx^i+\gb^idt)(dx^j+\gb^jdt)
\end{equation}
(in what follows $R\sim{}^3R$).
The momentum conjugate to to the metric is a density of weight one $\pi^{ij}=h^{1/2}
(Kh^{ij}-K^{ij})$ where $K_{ij}$ is the extrinsic curvature with trace K.  The natural
time derivative for evolution $\hat{\pp}_0$ acts in the normal future direction to the
spacelike slice $\gS$ and is denoted by an over-dot; one has $\hat{\pp}_0=\pp_t-{\mf
L}_{\gb}$ where ${\mf L}_{\gb}$ is the Lie derivative along the shift $\gb$.
Every foliation is described by a wave equation for N for some value of $\ga$ thus
making N a dynamical variable.  The Hamiltonian constraint does not fix the time but
does fix the proper time rate $d\tau/dt=\ga h^{1/2}=N$ along the normal $\pp_0$.
Using $\ga$ has the effect of altering the Hamiltonian density from H to
\bq\label{7.13}
\tl{H}=h^{1/2}H=\pi^{ij}\pi_{ij}-\frac{1}{2}\pi^2-hR
\end{equation}
which is of scalar weight 2 and a rational function of the metric.  $\tl{H}$ will be
referred to as the Hamiltonian density and may not vanish.  This leads to a modification
of the ADM action as in \cite{a6,t5}, namely ($16\pi G=c=1$)
\bq\label{7.14}
S(h,\pi,\ga,\gb)=\int d^4x(\pi^{ij}\dot{h}_{ij}-\ga\tl{H})
\end{equation}
(one assumes $N\sim 1+O(r^{-1})$).  Explicitly the Lie derivative term in
$\dot{\pi}^{ij}$ is, up to a divergence, $({\bf 7K})\,\,2\gb^i\na_j\pi^i_i=-\gb^iH_i$.
\\[3mm]\indent
Consider now a general variation of the modified Hamiltonian density
\bq\label{7.15}
\gd\tl{H}=(2\pi_{ij}-h_{ij}\pi)\gd\pi^{ij}+(2\pi^{ik}\pi_k^j-\pi\pi^{ij}+
\end{equation}
$$+h R^{ij}-hh^{ij}R)\gd h_{ij}-h(\na^i\na^j\gd h_{ij}-h^{ij}\na^k\na_k\gd h_{ij})$$
Note that this does not involve either the Hamiltonian or momentum densities; in contrast
the variation of the ADM Hamiltonian density $\gd H=\gd(h^{-1/2}\tl{H})$ does contain
a term proportional to the Hamiltonian density.  Requiring that S above be stationary
under a variation with respect to $\pi^{ij}$ gives the definition of the extrinsic
curvature
\bq\label{7.16}
\dot{h}_{ij}=\ga\frac{\gd\tl{H}}{\gd\pi^{ij}}=\ga(2\pi_{ij}-h_{ij}\pi)\equiv -2NK_{ij}
\end{equation}
Requiring stationarity under a variation in $h_{ij}$ gives the equation of motion
\bq\label{7.17}
\dot{\pi}^{ij}=-\ga\frac{\gd\tl{H}}{\gd h_{ij}}=-\ga
h(R^{ij}-h^{ij}R)-\ga(2\pi^{ik}\pi_k^j-
\end{equation}
$$-\pi\pi^{ij})+h(\na^j\na^i\ga-h^{ij}\na^k
\na_k\ga)$$
The slicing density $\ga$ and the shift $\gb^j$ are not to be varied; instead the
constraints are imposed on initial data and are preserved dynamically as shown below.
Thus consider the familiar 3+1 identities
\bq\label{7.18}
\dot{h}_{ij}\equiv -2NK_{ij};\,\,\dot{K}_{ij}\equiv N(R_{ij}-{}^4R_{ij}+KK_{ij}
-K_{ik}K^k_j-N^{-1}\na_i\na_jN)
\end{equation}
One recalls also that $h^{-1}\dot{h}=h^{ij}\dot{h}_{ij}=-2NK$.  Now pass to canonical
variables and use \eqref{7.18} to arrive at
\bq\label{7.19}
\dot{\pi}^{ij}\equiv Nh^{1/2}(Rh^{ij}-R^{ij})-Nh^{-1/2}(2\pi^{ik}\pi_k^j-\pi\pi^{ij})+
\end{equation}
$$+h^{1/2}(\na^i\na^jN-h^{ij}\na^k\na_kN)+Nh^{1/2}{\mf R}^{ij};\,\,{\mf R}_{ij}=
{}^4R_{ij}-h_{ij}{}^4R^k_k$$
\indent
One sees that the equations of motion (7.16)-(7.17) derived from the action principle 
are (7.18)-(7.19) when ${}^4R^{ij}-h^{ij}{}^4R_k^k=0$.  Thus to say that (7.17) holds
is to assert that ${}^4R^{ij}=0$.  In fact the equations of motion hold strongly 
independent of whether the constraints are satisfied or not and this is not true
in the ADM formulation because of the presence of the Hamiltonian density in the
equations of motion for $\pi^{ij}$.  This difference can be explained more fully as
follows.  Given $G_{\mu\nu}={}^4R_{\mu\nu}-(1/2)g_{\mu\nu}{}^4R^{\gs}_{\gs}$ and the
observation that $2G^0_0={}^4R^0_0-{}^4R^k_k$ one has $({\bf
7L})\,\,G_{ij}+h_{ij}G^0_0\equiv {}^4R_{ij}-h_{ij}{}^4R^k_k$.  The vanishing of the right
side does not depend on either the Hamiltonian or momentum densities and is equivalent to
${}^4R_{ij}=0$ or
$G_{ij}=-h_{ij}G^0_0$.  Thus while ${}^4R_{\mu\nu}=0$ and $G_{\mu\nu}=0$ are 
equivalent $R_{ij}=0$ and $G_{ij}=0$ are not equivalent as equations of motion - unless
the Hamiltonian density $H=2h^{1/2}G_0^0$ vanishes exactly (i.e. unless the Hamiltonian
constraint holds).  The ADM action principle is equivalent to $G_{ij}=0$ and one recalls
that the use of $R_{ij}$ instead of $G_{ij}$ has always been preferred by the French
school. This raises the important principle that a constrained Hamiltonian theory should
be well behaved even when the constraints are violated.  There is much further
calculation in this direction which we omit here (cf. \cite{p17,y2}).
\\[3mm]\indent
{\bf REMARK 7.1.}
There is a great deal of material now available on general relativity in terms of 
Ashtekar variables (see e.g. \cite{a6,a7,b7,g6,j1,k1,m2,r4,r6,s1,s14,s17,t2} for a very
incomplete list of references on loop quantum gravity, etc.).  In \cite{s17} for
example one recasts the WDW equation in the new variables in terms of the 3-geometry
elements C and ${\mf K}$ where C is the Chern-Simons functional and ${\mf K}$ is the
integral of the trace of the extrinsic curvature (cf. also \cite{s14}).$\hfill\bs$

\end{document}